\documentclass[english,aps,prb,superscriptaddress,]{revtex4-2}
\usepackage{babel}
\usepackage{amsmath,mathtools}
\usepackage{amssymb}
\usepackage{graphicx}
\usepackage{xcolor}
\def\Rafi{\textcolor{magenta}}
\begin{document}
\title{ Critical hybridization of skin modes in coupled non-Hermitian chains }
\author{S M Rafi-Ul-Islam }
\email{e0021595@u.nus.edu}
\selectlanguage{english}%
\affiliation{Department of Electrical and Computer Engineering, National University of Singapore, Singapore}
\author{Zhuo Bin Siu}
\email{elesiuz@nus.edu.sg}
\selectlanguage{english}%
\affiliation{Department of Electrical and Computer Engineering, National University of Singapore, Singapore}
\author{Haydar Sahin}
\email{sahinhaydar@u.nus.edu}
\selectlanguage{english}%
\affiliation{Department of Electrical and Computer Engineering, National University of Singapore, Singapore}
\author{Ching Hua Lee} 
\email{phylch@nus.edu.sg}
\selectlanguage{english}%
\affiliation{Department of Physics, National University of Singapore, Singapore}
\author{Mansoor B. A. Jalil}
\email{elembaj@nus.edu.sg}
\selectlanguage{english}%
\affiliation{Department of Electrical and Computer Engineering, National University of Singapore, Singapore}
\begin{abstract}
Non-Hermitian topological systems exhibit a plethora of unusual topological phenomena that are absent in the Hermitian systems. One of these key features is the extreme eigenstate localization of eigenstates, also known as non-Hermitian skin effect (NHSE), which occurs in open chains. However, many new and peculiar non-Hermitian characteristics of the eigenstates and eigenvlaues that emerge when two such non-Hermitian chains are coupled together remain largely unexplored. Here, we report  various new avenues of eigenstate localization in coupled non-Hermitian chains with dissimilar inverse skin lengths in which the NHSE can be switched on and off by the inter-chain coupling amplitude. A very small inter-chain strength causes the NHSE to be present at both ends of an anti-symmetric coupled system because of the weak hybridization of the eigenstates of the individual chains. The eigenspectrum under open boundary conditions (OBC) exhibits a discontinuous jump known as the critical NHSE (CNHSE) as its size increases. However, when the hybridization between eigenstates becomes significant in a system with strong inter-chain coupling, the NHSE and CNHSE vanish. Moreover, a peculiar `` half-half skin localization" occurs in composite chains with opposite signs of inverse decay lengths, where half of the eigenstates are exponentially localized at one chain and the remainder of the eigenstates on the other chain. Our results provide a new twist and insights for non-Hermitian phenomena in coupled non-Hermitian systems.      
\end{abstract}
\maketitle
\section*{Introduction}
Many intriguing fundamental physical and quantum phenomena, such as topological states \cite{lu2016topological,wang2017topological,rafi2020strain}, the quantum Hall effect \cite{cage2012quantum,bernevig2006quantum,sun2019field}, spin and valley Hall effect \cite{mak2014valley,tahir2013quantum,sun2020spin}, Berry curvature \cite{shimazaki2015generation,xiao2020berry} and quantum Klein and anti-Klein tunneling \cite{rafi2020anti, beenakker2008colloquium}, exist in condensed matter systems and their analogues that are described by Hermitian Hamiltonians. The ubiquitous properties of Hermitian systems, such as their mutually orthogonal eigenstates and real energy spectra, makes it possible to distinguish various topological states of matter based on various topological invariants (e.g., the Chern number \cite{tarnowski2019measuring,liu2012fractional,hofmann2019chiral}, winding numbers \cite{zhang2017direct,rafi2021topological}) derived from the eigenstates and eigenenergies.  The usual bulk-boundary correspondence (BBC), which describes the characteristics of topological boundary states \cite{meier2016observation,drozdov2014one},  is preserved in Hermitian systems.

However, most topological systems in reality experience a finite energy exchange from surroundings and an inevitable loss or gain of energy \cite{alvarez2018topological}. The non-conservation of energy results in non-Hermitian systems \cite{li2020topological,leykam2017edge,gong2018topological,li2020quantized,lee2020exceptional,lv2021curving,li2021non,zeuner2015observation,yokomizo2019non,borgnia2020non}, where the usual topological definitions and classifications of matter may no longer be applicable \cite{jin2019bulk,yao2018non}.   Interestingly, non-Hermitian systems host a plethora of unusual phenomena that include exceptional points \cite{kawabata2019classification,leykam2017edge,rafi2021non}, nodal rings \cite{wang2019non,bergholtz2021exceptional}, extensive localization of eigenstates \cite{hofmann2020reciprocal,lee2019hybrid,zhang2020non,kawabata2020higher}, unidirectional transport \cite{wu2018unidirectional,du2020controllable}, and the amplification and attenuation of  quantum signals \cite{roy2021nondissipative,el2015optical}.  To date, one of the most iconic features of a non-Hermitian system is the non-Hermitian skin effect (NHSE) \cite{okugawa2020second,rafi2021unconventional,shen2021non,kawabata2020higher,okuma2020topological,yi2020non,zhang2021tidal,lee2019anatomy,zhang2021universal,longhi2019probing, song2019non}, where the eigenstates experience extreme exponential localization in the vicinity of the boundaries.  Several non-Hermitian systems have been realized in various platforms including topolectrical \cite{xiao2020non,zou2021observation,ghatak2020observation,lee2018topolectrical,imhof2018topolectrical,rafi2020realization,rafi2020topoelectrical}, photonics \cite{feng2017non,midya2018non}, optics \cite{longhi2018parity,el2019dawn}, mechanical \cite{ghatak2020observation,schomerus2020nonreciprocal}, acoustic \cite{zhu2018simultaneous,shen2018synthetic} and superconductors \cite{ghatak2018theory,zhou2020non}, which offer broader accessibility and experimental flexibility in dynamic tuning of model parameters than condensed matter systems. 

To date, the unusual characteristics of one-dimensional non-Hermitian systems that have been theoretically or experimentally investigated are mostly for homogeneous systems consisting of a single chain. This motivates us to pose the question of whether new topological and skin effect phenomena may emerge if multiple dissimilar non-Hermitian chains are coupled together? The answer is affirmative. It has been recently reported that a new type of skin localization, known as the critical NHSE (CNHSE) \cite{li2020critical,liu2020helical,yokomizo2021scaling} occurs in a system consisting of two Nelson-Hatano chains, where the eigenenergy spectrum undergo a discontinuous transition across a critical point as the size of the system varies. A Nelson-Hatano chain is essentially a one-dimensional periodic chain with a unit-cell consisting of a single node in which the  hopping amplitude between a given node and its left neighbor differs from that between the node and its right neighbor. The non-reciprocity in the direction results in non-Hermiticity in the system, and can be characterized by the inverse skin length of the resultant exponentially localized skin modes. When the chains are short and the inter-chain coupling is weak, the energy spectrum of the coupled chains resembles that of the individual uncoupled chains and take the form of lines parallel to the real or imaginary axis on the complex energy plane (i.e., the short-chain regime). When the length of the chains exceeds a 
critical length, the energy spectrum, rather surprisingly, undergoes a qualitative transition, i.e., the CNHSE, in which they transit from lines to closed curves enclosing finite areas on the complex energy plane (i.e., the long-chain regime). 

In this work, we further investigate how the interplay between the inter-chain coupling amplitude and the non-Hermitian parameters results in the different types of skin mode localization, including the CNHSE, in a pair of coupled non-Hermitian chains with dissimilar inverse skin lengths. We find that the emergence of the CNHSE is contingent on the strength of the inter-chain coupling between the chains. A stronger inter-chain coupling pushes the length at which the CNHSE occurs and the system transits from the short-chain regime to the long-chain regime to shorter lengths, until eventually the short-chain regime vanishes completely for sufficiently strong inter-chain coupling. We explain the CNHSE and its dependence on the inter-chain coupling strength, as well as the subsequent behavior of the eigenenergy spectrum in the long-chain regime as the length of the system increases further beyond the CNHSE critical length. 

In conjunction with the effect of the inter-chain coupling on the CNHSE, we investigate how the relative values of the non-Hermitian parameters influence the localization of the skin modes. We show that the eigenstates of the coupled chains undergo constructive (destructive) hybridization, when the individual chains have the same (opposite) signs of the inverse decay lengths. The eigenstates will all pile up exponentially either near the left- or right-most sites under OBC for constructive hybridization. In contrast, for destructive interference, a peculiar type of localization occurs where a portion of the eigenstates are exponentially localized at the right-most sites of one chain, and the remainder at the left-most sites of the other chain. We dubbed this as `` half-half skin localization''. 
  
\section{Model}
\subsection{Model and characterization of a coupled chain system}
\begin{figure}[ht!]
\centering
\includegraphics[width=0.7\textwidth]{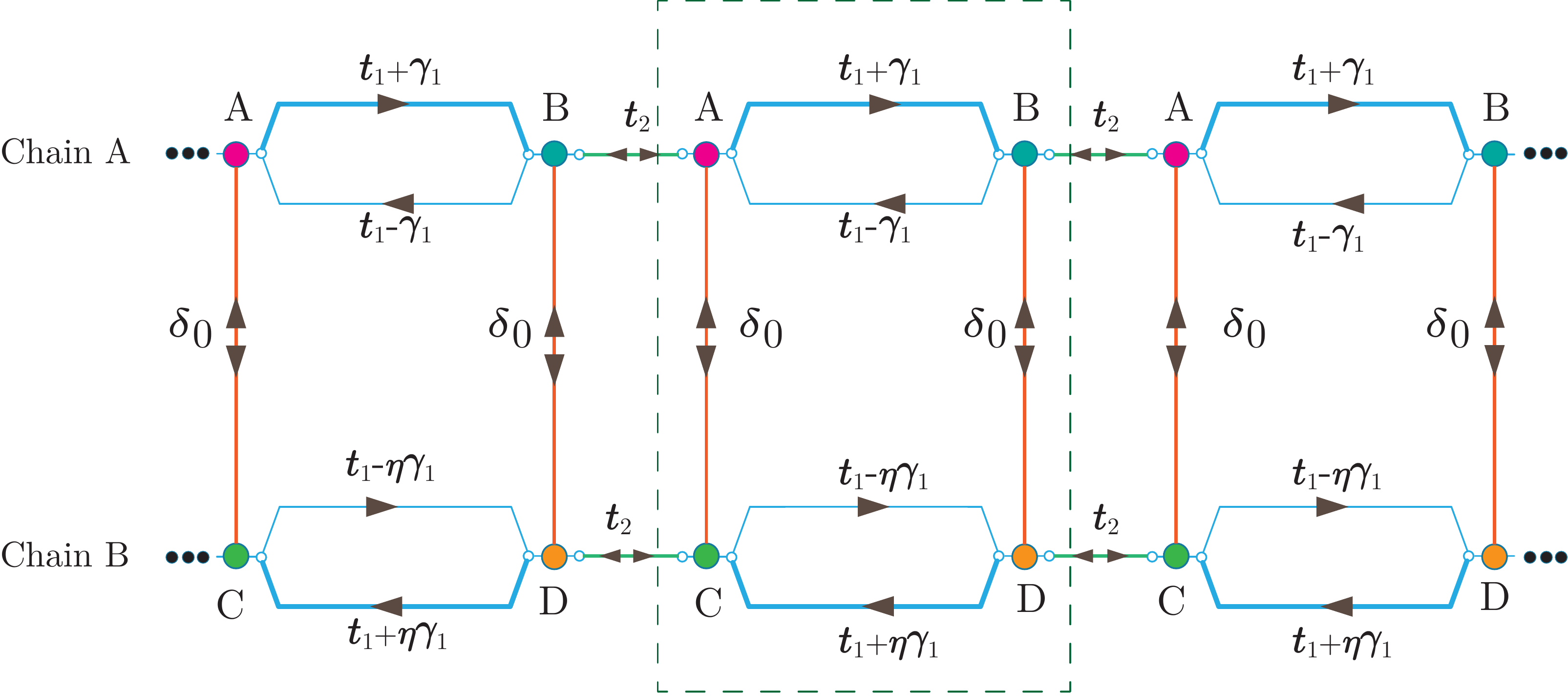}
\caption{ \textbf{Schematic representation of a pair of coupled non-Hermitian chains.} The imbalance in the forward and backward coupling that is induced by the $\gamma_1$ and  $\eta \gamma_1$ terms in chains A and B, respectively, gives rise to non-Hermiticity  and extreme eigenstate localization in the isolated chains. The unit cell is marked by the dashed rectangle, within which the intra-cell coupling ($t_2$) is reciprocal while the intra-cell coupling contains hopping asymmetry. The inter-chain coupling $\delta_0$ connects two chains, and causes the mixing of the eigenstates of chains A and B.}
\label{gFig1}
\end{figure}

We consider a simple pair of two coupled non-Hermitian SSH chains with dissimilar inverse decay lengths (see Fig. \ref{gFig1}) where within each chain, the intra-unit cell and inter-unit cell couplings are respectively non-reciprocal and reciprocal in nature. There are various platforms, such as photonic, optical, and topolectrical systems, through which such coupled systems can be realized \cite{weidemann2020topological,zou2021observation,stegmaier2021topological,helbig2020generalized,koh2021stabilizing,li2020critical,liu2019topological,esmann2018topological}. We emphasize that the results and characteristics of our coupled model are independent of the choice of the implementation platform so that the most easily accessible platform can be chosen for the experimental realization.     
We consider a reciprocal but tunable coupling ($\delta_{AB}=\delta_{BA}= \delta_0$) that couples the two non-Hermitian SSH chains (see Fig. \ref{gFig1}). The Hamiltonian of the coupled chains for Fig. \ref{gFig1} can be expressed as 
 \begin{equation}
 H_{\mathrm{two-chains}} (k_x)= \begin{pmatrix}
(t_1+t_2 \cos k_x) \sigma_x +(t_2 \sin k_x -i \gamma_1) \sigma_y & \delta_0 \mathbf{I}_{2\times 2} \\
\delta_0 \mathbf{I}_{2\times 2} & (t_1+t_2 \cos k_x) \sigma_x +(t_2 \sin k_x +i \eta \gamma_1) \sigma_y
\end{pmatrix},
\label{reeq1}
\end{equation}
where the two diagonal terms in Eq. \ref{reeq1} describe the Hamiltonian for the isolated chains and $\sigma_i$ and $\mathcal{I}_{2\times 2}$ are the $i$th Pauli matrix and identity matrix respectively. The terms $\gamma_1$ and $\eta \gamma_1$  induce non-Hermiticity in the system via creating an imbalance between the forward/backward coupling for chains A and B, respectively.

We note for later reference that the hybridization between the eigenstates of the individual chains in the coupled system is hinted at the Schr\"{o}dginer equation for the Hamiltonian Eq. \eqref{reeq1}. Consider a generic Hamiltonian $H_{\mathrm{coupled}}$ for a coupled system comprising two subsystems:
\begin{equation}
	H_{\mathrm{coupled}} = \begin{pmatrix} H_{\mathrm{A}} & \delta_0 \\ \delta_0 & H_{\mathrm{B}} \end{pmatrix} \label{Hcoupled} 
\end{equation} 
where the $H_{\mathrm{A, B}}$s are the Hamiltonians of the two isolated systems, and $\delta_0$ is the coupling between them. Writing the eigenstate of $H_{\mathrm{coupled}}$ as $|\Psi\rangle = (|\psi_{\mathrm{A}} \rangle, |\psi_{\mathrm{B}}\rangle)^{\mathrm{T}}$ and the eigenenergy of $|\Psi\rangle$, expanding the time-independent Schr\"{o}dinger equation $H_{\mathrm{coupled}}|\Psi\rangle = |\Psi\rangle E$ gives the equations 
\begin{align} 
	H_{\mathrm{A}} |\psi_{\mathrm{A}}\rangle + \delta_0|\psi_{\mathrm{B}}\rangle &= |\psi_{\mathrm{A}}\rangle E, \label{tiseHcoupled1}   \\ 
	H_{\mathrm{B}} |\psi_{\mathrm{B}}\rangle + \delta_0|\psi_{\mathrm{A}}\rangle &= |\psi_{\mathrm{B}}\rangle E. \label{tiseHcoupled2}
\end{align}  
Eq. \ref{reeq1} has the generic form of Eq. \eqref{Hcoupled} where $H_{\mathrm{A}}$ and $H_{\mathrm{B}}$ can be identified with the Hamiltonains of the two uncoupled chains. The presence of both $|\psi_{\mathrm{A}}\rangle$ and $|\psi_{\mathrm{B}}\rangle$ in Eqs. \eqref{tiseHcoupled1} and \eqref{tiseHcoupled2} due to the coupling term $\delta_0$ in Eq. \eqref{Hcoupled} demonstrates the mixing of the contributions from both chains in the eigenstates of the coupled chain and the role that $\delta_{\mathrm{AB}}$ and $\delta_{\mathrm{BA}}$ formally play in determining the degree of mixing between the two subsystems. Therefore, the wavefunction distribution of a pair of coupled one-dimensional chains shows a high sensitivity to the inter-chain coupling strength and may possess very different characteristics compared to the isolated chains, as we shall now proceed to show in the sections that follow. 
  
\section{Results and discussion}
\begin{figure}[ht!]
\includegraphics[width=0.7\textwidth]{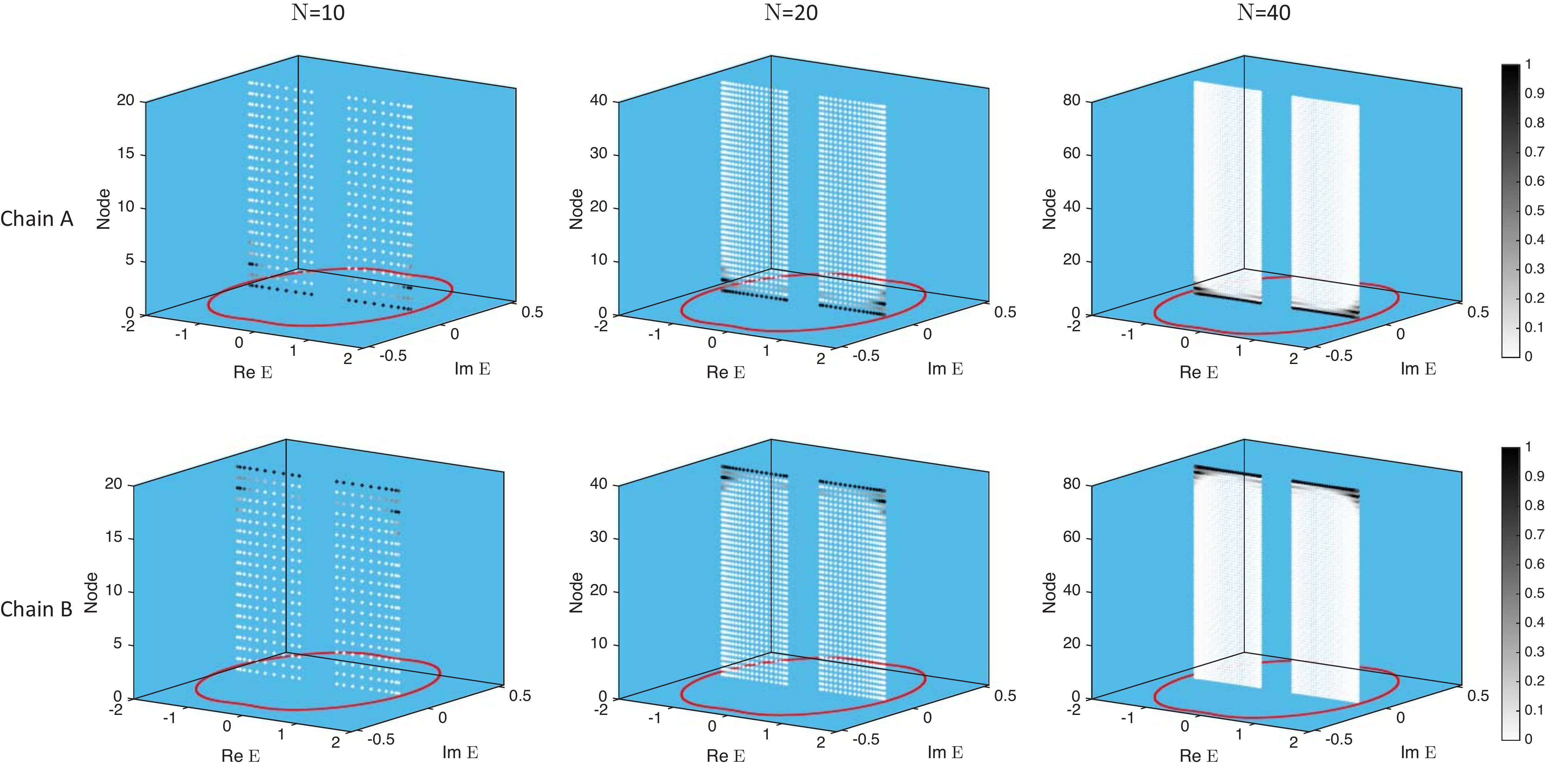}
\caption{\textbf{Size-independent eigenvalues and eigenstate distributions of isolated chains.} (a) A and (b) B, where $\eta=1$, with different finite lengths of $N=$ 10, 20, and 40 unit cells. The red dots represent the PBC eigenenergy spectra of the chains. The $(x,y)$ coordinates of the dots indicate the complex energy values of the eigenstates. For each eigenstate, the $z$ coordinate of each dot indicates the spatial positions of a lattice point in the system and the color of each dot indicates the relative magnitude of the square of the wavefunction with darker colors indicating larger magnitudes.  Common parameters: $t_1=1.2$, $t_2= 0.6$ and $\gamma_1= 0.8$. This plot is just to connect with familiar uncoupled chains by plotting their eigenspectrum and eigenstates, which are size-independent. }
\label{gFig2}
\end{figure}

To establish the ground for further discussion, we first describe the non-Hermitian skin effect of the isolated chains. Fig. \ref{gFig2} shows the eigenenegy spectra and eigenstate density distributions of the isolated chains of various lengths. This figure serves as a reference for comparing the various size-dependent non-Hermitian characteristics that emerge when the chains are coupled later. Under OBC and in the thermodynamic limit, the properties of the resulting system are described by the surrogate Hamiltonian in which the replacement $\exp(ik)\rightarrow\beta$ is made, and $k$ is, in general, complex. In this limit, the isolated chains $A$ and $B$ exhibit the non-Hermitian skin effect (NHSE) with the inverse decay lengths of $\kappa_a = \frac{1}{2} \log (\frac{t_1-\gamma_1}{t_1+\gamma_1})$ and  $\kappa_b = \frac{1}{2} \log (\frac{t_1+ \eta \gamma_1}{t_1- \eta \gamma_1})$ where the inverse skin length $\kappa$ is given by $\kappa=\ln |\beta|$ . For positive values of $t_1$, $\gamma_1$ and $\eta$, the non-Hermitian skin modes are localized near the left-most and right-most nodes for chain $A$ and chain $B$, respectively (see Fig. \ref{gFig2}). Furthermore, the OBC eigenvalue distribution of the isolated chains, which takes the form of two separated lines on the real axis symmetrically distributed about the imaginary axis, does not vary with the system size $N$ (see first and third columns of Fig. \ref{gFig2}).

To help explain the effect of the inter-chain coupling $\delta_0$ on the skin modes distribution of the coupled chain when we present the results for each scenario later, we write down for later reference the characteristic equation of the surrogate Hamiltonian Eq. \eqref{reeq1} as 
\begin{equation}
\delta_0^4 + \Delta (\beta, E) \delta_0^2 +  f_{\mathrm{A}}(\beta, E)f_{\mathrm{B}}(\beta, E)\delta_0^0 = 0,
\label{reeq3}
\end{equation}
where $\beta =  e^{i(k+i \kappa)}$ is the non-Bloch factor with complex momentum $k+i\kappa$, $\Delta (\beta, E)=2(\eta \gamma_1^2 -t_1^2 - t_2^2 - E^2)+\beta^{-1}((t_2 \gamma_1 (1+ \eta)(\beta^2-1)-(1+ \beta^2)2 t_1 t_2))$ and $f_i (\beta, E)=E^2-\beta^{-1}(t_2+\beta (t_1-\zeta_i \gamma_1)(t_1+t_2 \beta + \zeta_i \gamma_1))$, where $\zeta_1 =+1$ and $\zeta_1 =- \eta$ for chains A and B respectively. Interestingly, $f_i (\beta, E)$ is the characteristic polynomial of chain $i$. The coefficients of $\delta_0^4$ and $\delta_0^2$ indicate the degree of hybridization between the eigenstates of the two chains. 
\begin{figure}[ht!]
\centering
\includegraphics[width=0.7\textwidth]{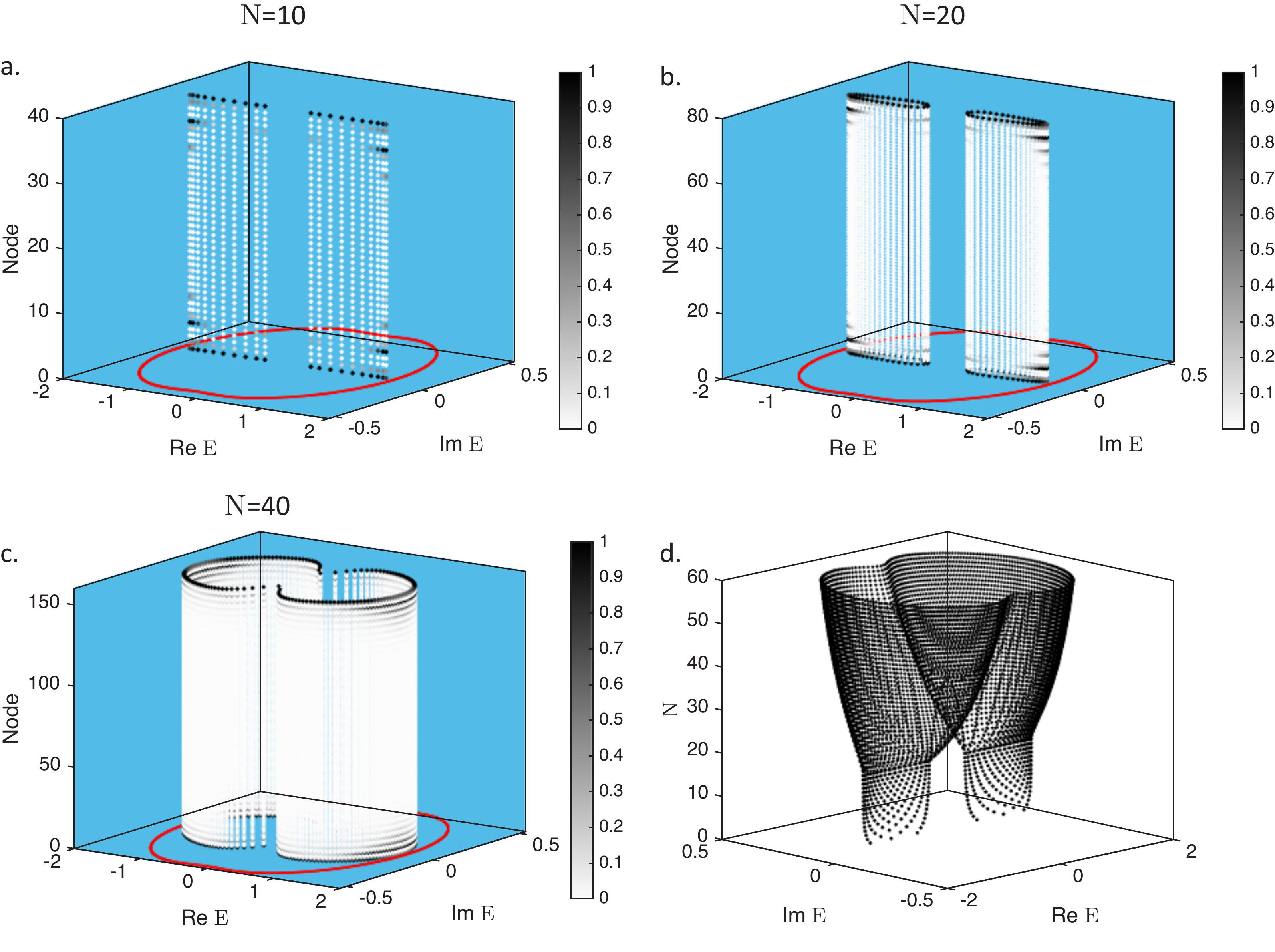}
\caption{\textbf{Size-dependence of CNHSE, NHSE, mode localization and eigenvalue dispersion, under OBC for nearly decoupled chains with $\delta_0 = 10^{-7}$.} The red dots represent the PBC eigenenergy spectra of the chains. The $(x,y)$ coordinates of the dots indicate the complex energy values of the eigenstates. For each eigenstate, the $z$ coordinate of each dot indicates the spatial positions of a lattice point in the system. The color of each dot indicates the relative magnitude of the square of the wavefunction with darker colors indicating larger magnitudes.  \textbf{a, b, c} Eigenvalue and eigenstate distribution of Eq. \ref{reeq1} at the different system sizes of $N=10$, $20$, and $40$ unit-cells for the inter-chain coupling. Clearly, the skin modes are localized at the left-most and right-most sublattice sites of chains A (smallest $z$ coordinate value) and B (largest $z$ coordinate value), respectively. The OBC spectra is real-valued up to small $N = 20$, reflecting the negligible effect of $\delta_0$. However, the OBC spectra expands into complex plane at larger system sizes and gradually approaches towards the PBC spectra as the system size increases. \textbf{d} shows the variation of the eigenvalue distribution with the system size. Common parameters used: $t_1=1.2$,  $t_2= 0.6$, $\gamma_1= 0.8$, $\eta=1$, and $\delta_0= 1 \times 10^{-7}$. Note the shift in the shift from real to complex eigenvalues as the system size ($N$) increases, which signifies the occurrence of the CNHSE.    }
\label{gFig3}
\end{figure}

To elucidate the effect of non-zero $\delta_0$  on the eigenstate and eigenvalue distribution under OBC for the coupled chains, we first consider the scenario of $\eta=1$, at which the forward and backward hoppings in chain $A$ are respectively equal to the backward and forward hoppings in chain $B$, and consequently, $\kappa_a=-\kappa_b$. The resultant skin mode localization shows a high sensitivity  to the inter-chain coupling $C_c$, as detailed below. For a very small inter-chain coupling magnitude  (e.g, $\delta_0=10^{-7}$ in the nearly decoupled case in Fig. \ref{gFig3} ), the higher order terms in $\delta_0$ in Eq. \ref{reeq3} can be neglected. To a good approximation, Eq. \eqref{reeq3} takes the form of the product of the characteristic equations of the individual chains. Therefore, skin modes solutions exist for both chains A and B and the non-Hermitian skin modes are localized at the both ends of the chains (see Fig. \ref{gFig3}a--c). The OBC spectra take the same form of two separated lines on the real axis as the isolated chains for small values of $N$ (see Fig. \ref{gFig3}a). When the system length exceeds a threshold value, the OBC spectrum gains an imaginary part, and slowly extends towards the form of the PBC spectrum at larger system sizes (Fig. \ref{gFig3}b--d). The transition of the OBC spectra from the isolated-chain form of two lines on the real axis to the form of a closed lozenge-shaped curve on the complex plane occurs at $N=20$. This switch in the form of the OBC spectrum with the system size is a manifestation of the CNHSE, which we explain in the Appendix. 
   
\begin{figure}[ht!]
\centering
\includegraphics[width=0.7\textwidth]{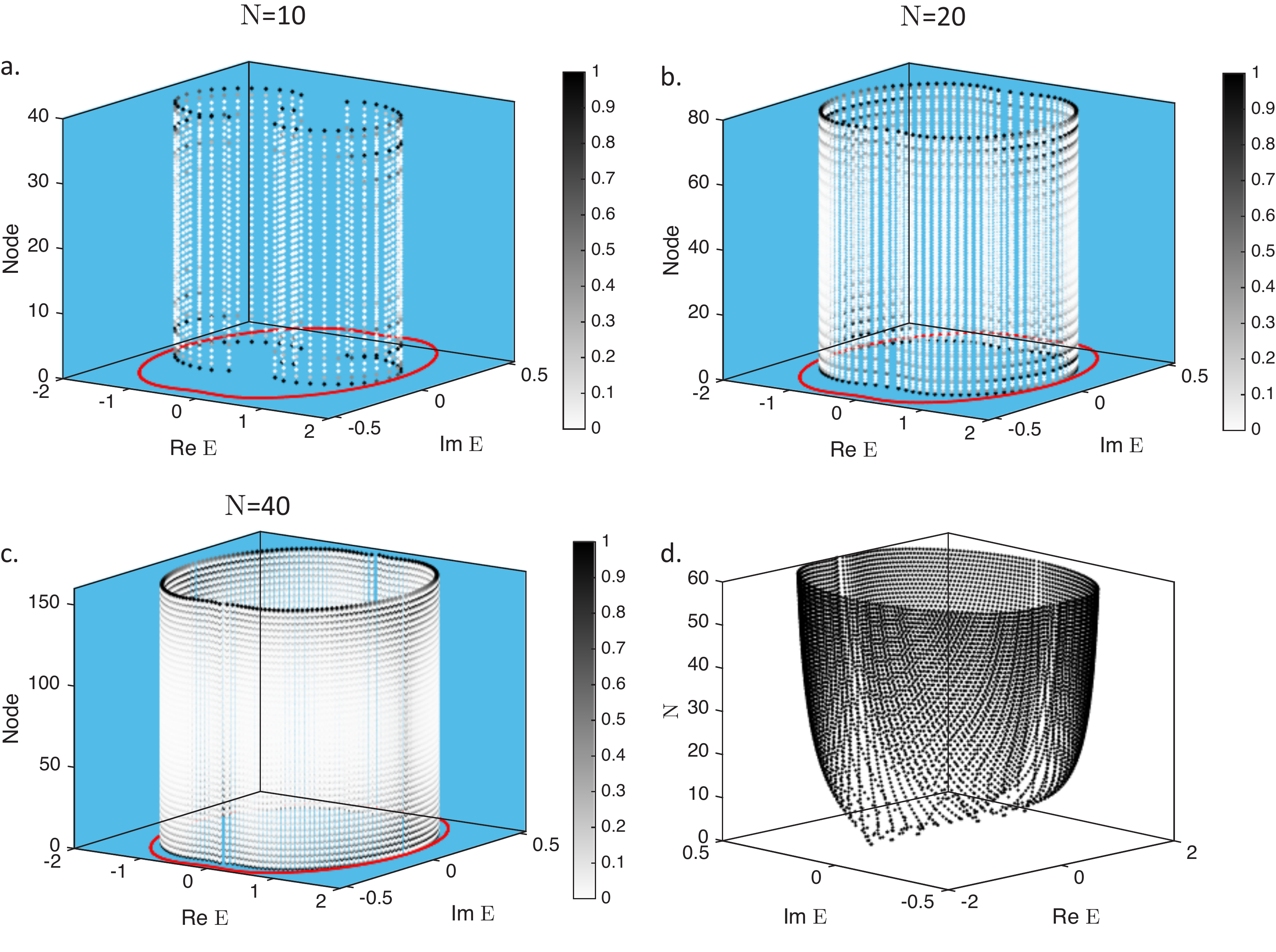}
\caption{\textbf{Destructive hybridization of skin modes in a weakly coupled antisymmetric chain system. } NHSE mode localization and eigenvalue dispersion under OBC for the weakly decoupled chains with $\delta_0 = 5\times 10^{-3}$. The red dots represent the PBC eigenenergy spectra of the chains. The $(x,y)$ coordinates of the dots indicate the complex energy values of the eigenstates. For each eigenstate, the $z$ coordinate of each dot indicates the spatial positions of a lattice point in the system. The color of each dot indicates the relative magnitude of the square of the wavefunction with darker colors indicating larger magnitudes. \textbf{a, b, c} Eigenvalue and eigenstate distribution of Eq. \ref{reeq1} at the different system sizes of $N=10$, $20$,  and $40$ unit-cells for the inter-chain. Clearly, the skin modes are localized at the left-most and right-most sublattice sites of chains A (smallest $z$ coordinate value) and B (largest $z$ coordinate value), respectively. The OBC spectra is real-valued up to small $N = 5$, reflecting the negligible effect of $\delta_0$. However, the OBC spectra expands into the complex plane at larger system sizes and gradually approaches towards the PBC spectra as the system size increases. \textbf{d} shows the variation of the eigenvalue distribution with the system size. Common parameters used $t_1=1.2$,  $t_2=0.6$, $\gamma_1= 0.8$, $\eta=1$, and $\delta_0= 5 \times 10^{-3}$. }
\label{gFig4}
\end{figure}

When $\delta_0$ is small but not negligible (i.e., $\delta_0=0.005$ ), the two chains are weakly coupled and the higher order $\delta_0$ terms in Eq. \eqref{reeq3}  cannot be neglected. As a result, the skin modes solution for the coupled system is no longer well approximated as the superposition of the skin mode solutions of the individual chains. Rather, the skin modes contributions of chains A and B become hybridized. We call this hybridization a destructive hybridization because $\kappa_a$ and $\kappa_b$ have opposite signs. The skin modes are still localized at the two ends but with a smaller magnitude of the localization for a given system size $N$ compared to the nearly decoupled case (see Fig. \ref{gFig4}a--c) case because of the larger hybridization due to the larger magnitude of $\delta_0$, which determines the mixing  of the eigenstates from two individual chains. Furthermore, the CNHSE occurs at a smaller value of $N$, and the OBC spectra also tends to approach the PBC spectra faster with the increase in $N$ (see Fig. \ref{gFig4}d).
 
\begin{figure}[ht!]
\centering
\includegraphics[width=0.7\textwidth]{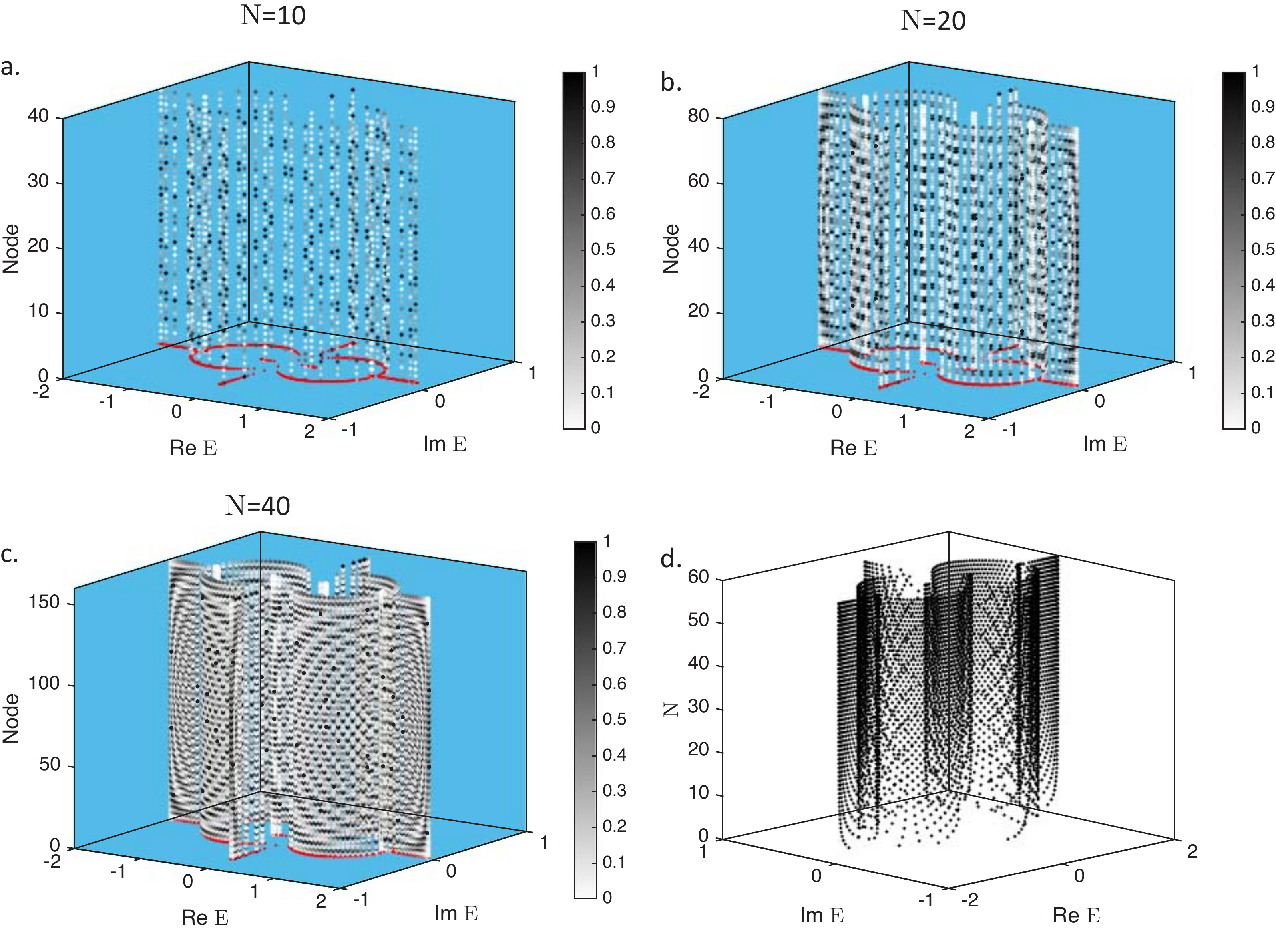}
\caption{\textbf{Vanishing NHSE and CNHSE induced by strong inter-chain coupling in a coupled antisymmetic chain system.} NHSE mode localization and eigenvalue dispersion under OBC for the strongly coupled chains with $\delta_0 = 0.3$. The red dots represent the PBC eigenenergy spectra of the chains. The $(x,y)$ coordinates of the dots indicate the complex energy values of the eigenstates. For each eigenstate, the $z$ coordinate of each dot indicates the spatial positions of a lattice point in the system. The color of each dot indicates the relative magnitude of the square of the wavefunction with darker colors indicating larger magnitudes.  \textbf{a, b, c} Eigenvalue and eigenstate distribution of Eq. \ref{reeq1} at the different system sizes of $N=10$, $20$, and $40$ unit-cells. The skin mode localization vanishes owing to $\kappa_a =-\kappa_b$ and the large $\delta_0$. The OBC and PBC spectra are complex valued and coincide with each other for at all the system sizes shown, indicating the absence of  size-dependent dispersion and the dominating effect of $\delta_0$. \textbf{d} shows the variation of the eigenvalue distribution with the system size. Common parameters used: $t_1=1.2$,  $t_2= 0.6$, $\gamma_1= 0.8$, $\eta=1$, and $\delta_0= 0.3$.  }
\label{gFig5}
\end{figure} 

When the inter-chain coupling $\delta_0$ is large enough (i.e., $\delta_0 \approx 0.3$ in Fig. \ref{gFig5}), the contributions of the first term in Eq. \eqref{reeq3} dominate over other terms. As  a result, the two chains are strongly coupled and the eigenmodes from the two chains are extensively hybridized. Because $\kappa_a= -\kappa_b$, the skin modes from the chains are hybridized destructively, and the NHSE for the coupled systems vanishes even at a small system size.   Furthermore, no critical transition can be found in the skin modes of the strongly coupled system.  In contrast to the weakly coupled case of $\delta_0 = 0.005$ where the switch-over occurs at the small values of $N=5$ to $N=6$, there is no discernable occurrence of the isolated dispersion relations in the strong coupling case of $\delta_0=0.3$. The area occupied by the OBC dispersion spectra on the complex energy plane increases with $N$ until the OBC dispersion spectra approach the thermodynamic limit OBC spectra. 

\subsection{Destructive hybridization of skin modes}
\begin{figure}[ht!]
\centering
\includegraphics[width=0.7\textwidth]{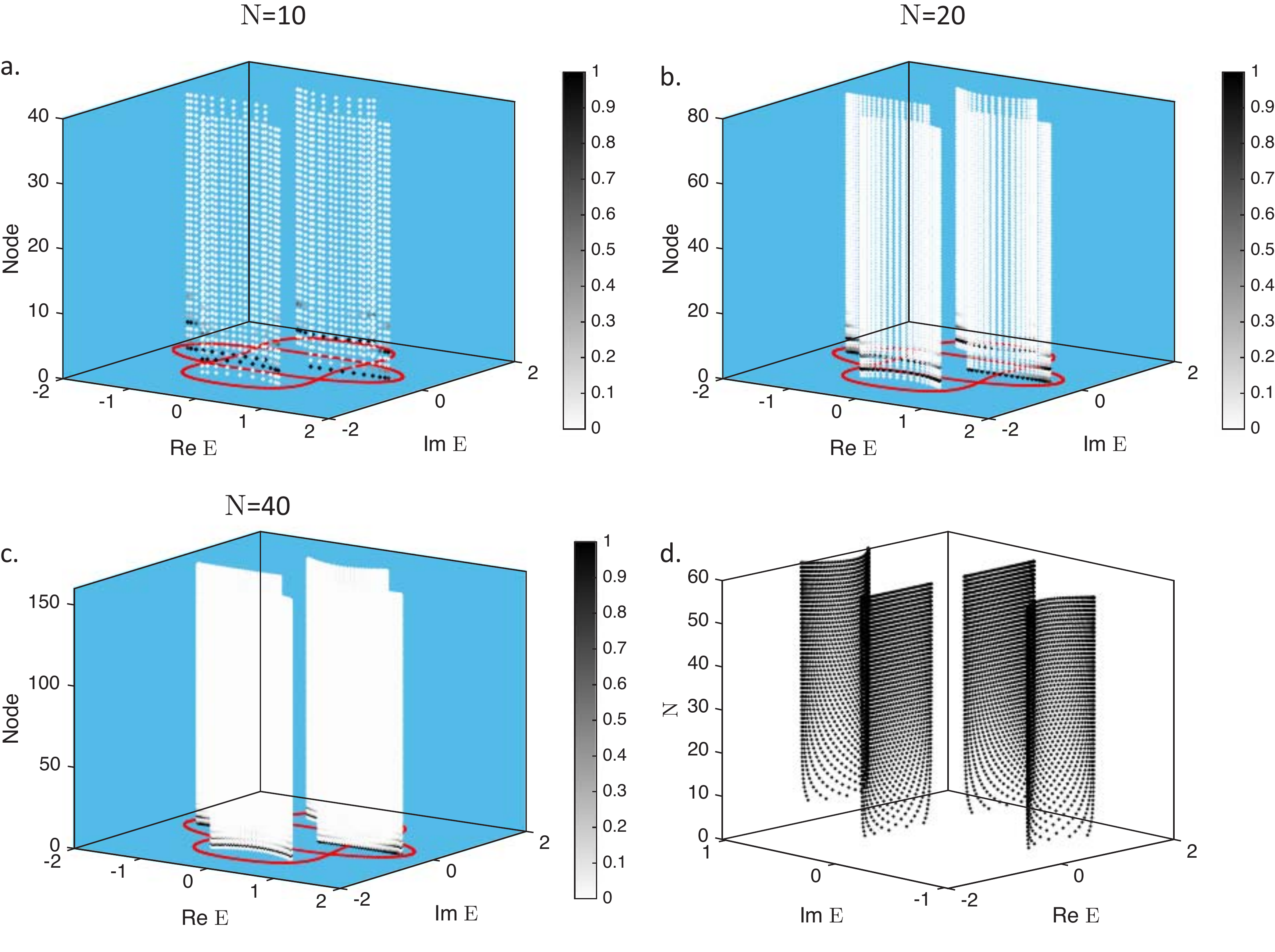}
\caption{\textbf{Constructive hybridization of the skin eigenmodes for finite chains with \textbf{a, b, c} 10, 20, and 40 unit cells, respectively.} Both chains possess the same signs of the inverse decay lengths $\kappa_a$ and $\kappa_b$. The red dots represent the PBC eigenenergy spectra of the chains. The $(x,y)$ coordinates of the dots indicate the complex energy values of the eigenstates. For each eigenstate, the $z$ coordinate of each dot indicates the spatial positions of a lattice point in the system. The color of each dot indicates the relative magnitude of the square of the wavefunction with darker colors indicating larger magnitudes. Half of the eigenstates are localized at the left-most site of the chain A , and the remainder at the left-most site of chain B under OBC. \textbf{d} shows the variation of the eigenvalue distribution with the system size. Common parameters used: $t_1=1.2$,  $t_2= 0.6$, $\gamma_1= 0.8$,  $\eta= -2$ and $\delta_0 = 0.2$ }
\label{gFig6}
\end{figure}
The signs of the inverse decay lengths of the individual chains affect the nature of hybridization between eigenstates for a coupled system. We have so far considered the case of $\eta=1$. If $\eta$ now takes a negative value, the forward hoppings in both chains will be stronger than their respective backward hoppings. As a result, $\kappa_a$ and $\kappa_b$ have the same sign if $|\eta \gamma_1|< |t_1|$. Therefore, the skin modes from chains A and B are hybridized constructively and the resultant skin modes are localized at either the left or right end nodes of the systems (see Fig. \ref{gFig6}a--c). In particular, when $\eta=-1$ so that $\kappa_a = \kappa_b$ and $\kappa_a < 0$, half of the eigenstates will be localized at the left-most unit cell of chain A (nodes 1 and 2 in the figure)  and the other half will be accumulated at the left-most unit cell of the chain B reflectivity (nodes 3 and 4 on the plots). Interestingly, the OBC spectra remains the same with the variation of the system size $N$, indicating the absence of the the critical skin effects. 
 
\begin{figure}[ht!]
\centering
\includegraphics[width=0.7\textwidth]{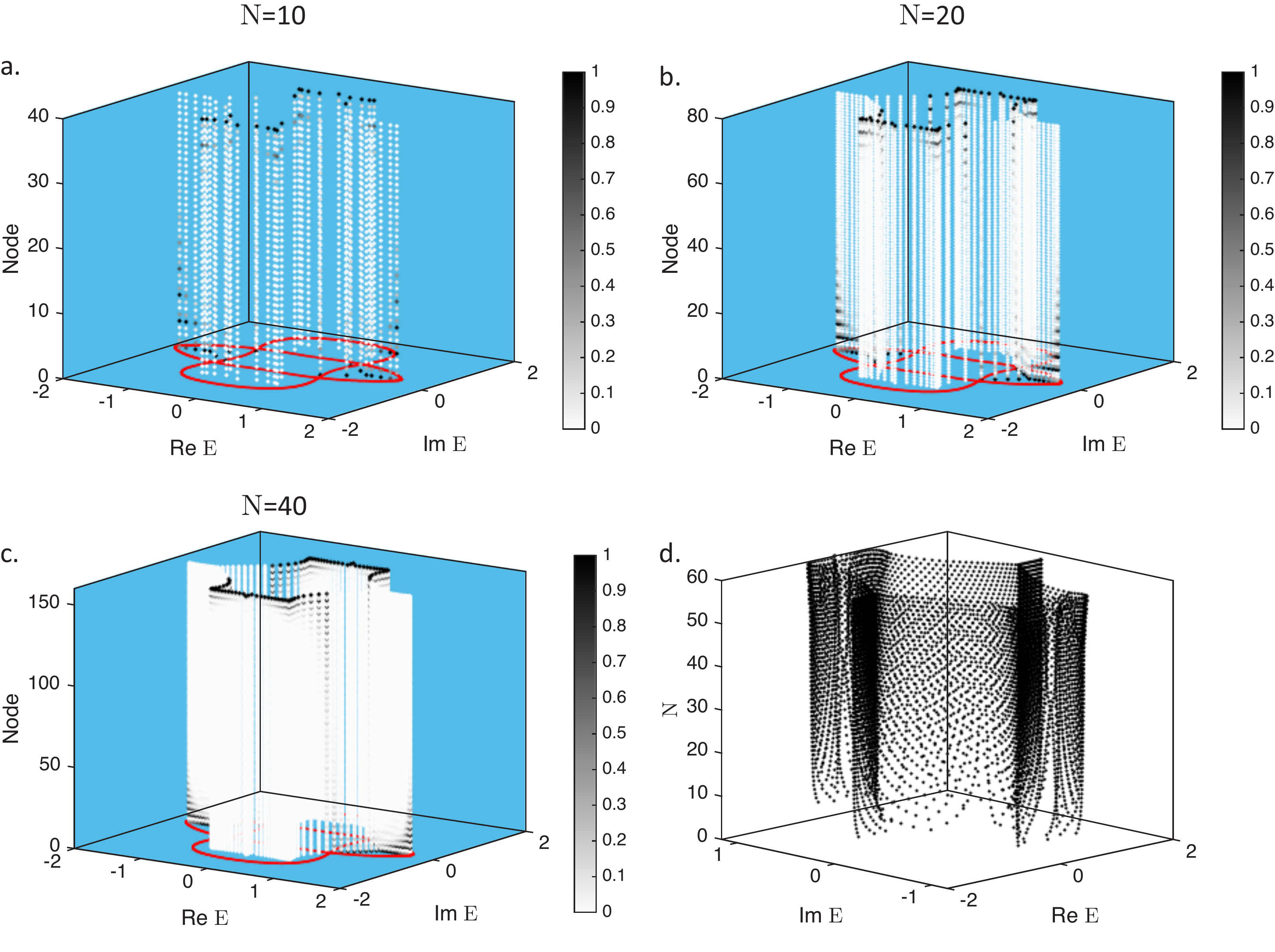}
\caption{\textbf{Demonstration of partially destructive hybridization of the skin eigenmodes.} The coupled system consists of chains with  pposite signs of inverse decay lengths $\kappa_a, \kappa_b$. \textbf{a, b, c} NHSE mode localization at different system size $N=10$, $20$, and $40$ unit-cells. Clearly, half of the eigenstates are localized at the left-most site of chain A, and the remainder at the right most site of chain B, respectively under OBC. We dub this as ``half-half skin localization''. \textbf{d} shows the variation of the eigenvalue distribution with the system size.  Common parameters used: $t_1=1.2$,  $t_2= 06$, $\gamma_1= 0.8$.
}
\label{gFig7}
\end{figure}

In contrast, if the chains in the coupled system possess inverse decay lengths with opposite signs,  eigenstates will experience destructive hybridization. Figs. \ref{gFig3} -- \ref{gFig5} show examples of perfectly destructive at which $\eta=1$. We now consider partially destructive hybridization where $\eta > 0$ and $\eta \neq 1$. In this case, $\kappa_a$ and $\kappa_b$ will exhibit mutually opposite signs when $|\eta \gamma_1|< |t_1|$, at which there is  partially destructive hybridization between the skin modes in the chains with $\kappa_a \neq -\kappa_b \neq 0$. The skin modes from two chains will undergo partially destructive interference. Some of the eigenstates of the coupled TE chains are localized near the left-most node of the chain A and the others near the right-most node of the chain B. We dub this peculiar skin modes localization as the ``half-half skin localization" (see Fig. \ref{gFig7}a--c) For a large range of $\eta$ values, including the one shown in Fig. \ref{gFig7}, exactly half of the eigenstates are localized near the left-most node and half near the right-most node (see the Appendix). 

\begin{figure}[ht!]
\centering
\includegraphics[width=0.7\textwidth]{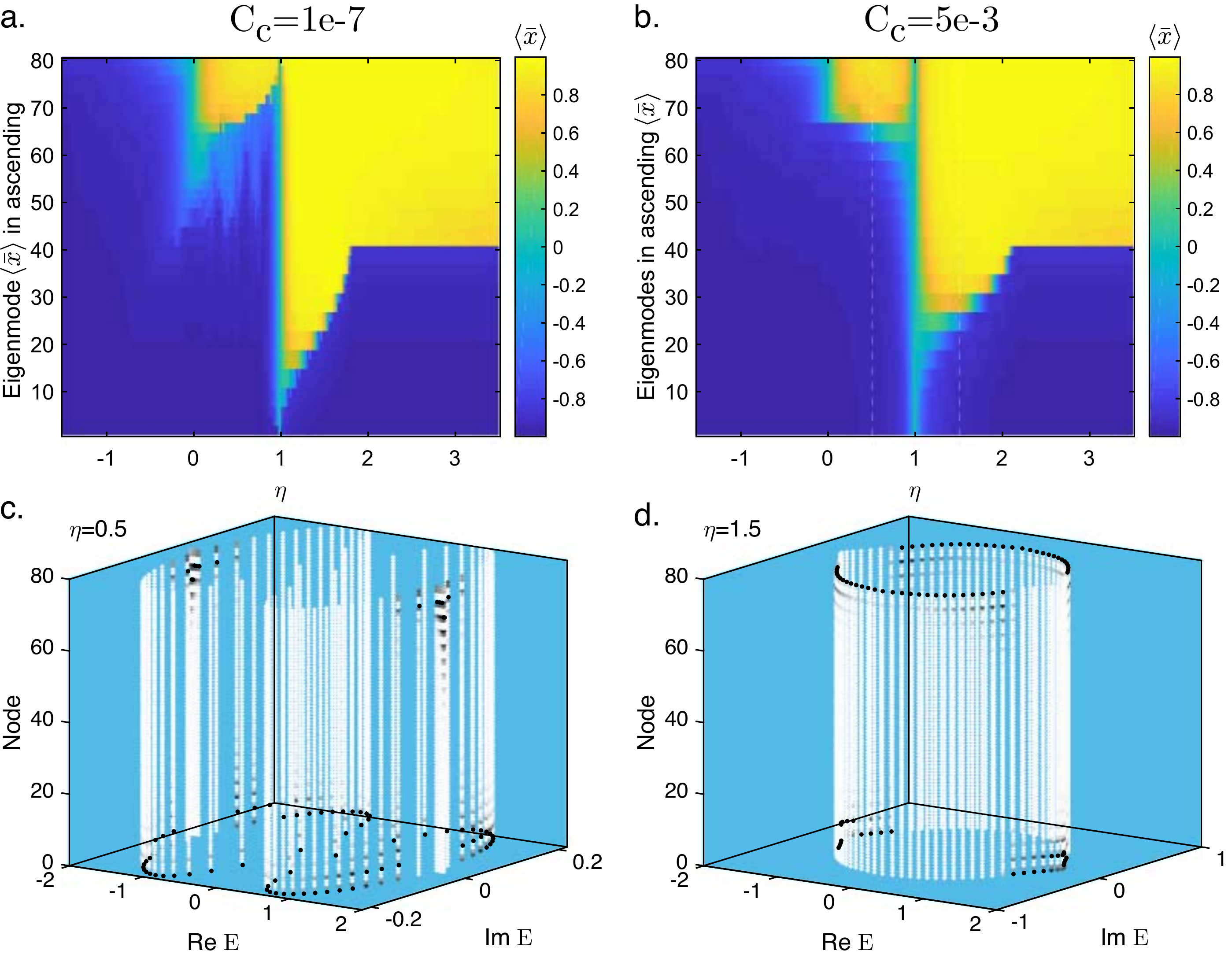}
\caption{\textbf{The preferred edges for skin modes localization as function of the ratio of the non-reciprocity in the two chains $\eta$ for different values of the interchain-coupling $\delta_0$, and exemplary spatial mode distributions.}  \textbf{a}, \textbf{b} show the normalized position expectation values of the eigenmodes $\langle \bar{x} \rangle$ (see text) sorted in increasing order of $\langle \bar{x} \rangle$ as functions of $\eta$ at \textbf{a} $\delta_0 = 10^{-7}$, and \textbf{b} $\delta_0 = 0.005$. \textbf{c} and \textbf{d} shows the spatial and energy distributions of the $\delta_0=0.005$ eigenstates at \textbf{c} $\eta=0.5$ and \textbf{d} $\eta=1.5$. These two values of $\eta$ are indicated by the vertical white dotted lines in panel b.  Common parameters used: $t_1=1.2$,  $t_2= 06$, $\gamma_1= 0.8$,  and $N = 20$  }
\label{gFig9}
\end{figure}

We further investigate how the localization of the eigenstates change under partially destructive interference with the variation of $\eta$, the ratio between the non-reciprocal couplings between the two chains. To quantify the localization of the eigenstates, we calculate the `normalized position expectation value' $\langle \bar{x} \rangle$ defined as $\langle \bar{x} \rangle = \sum_{i, \alpha} \bar{x}_i |\psi_{i\alpha}|^2$ where the summation index $i$ indicates the unit cell, the index $\alpha$ denotes the $\alpha$th node within the unit cell, and the normalized position $\bar{x}_1 = -1$ at the left-most unit cell and $\bar{x}_N = 1$ at the right-most unit cell.  A $\langle\bar{x}\rangle$ value of $1$ ($-1$) therefore indicates that the mode is localized completely at the right-most (left-most) unit cell.

The variation of the eigenstate localization as indicated by the normalized position with $\eta$ for two exemplary values of $\delta_0$ are shown in Fig. \ref{gFig9}. Fig. \ref{gFig9}a,b shows the normalized position expectation values of all the eigenmodes arranged in order of ascending $\langle\bar{x}\rangle$ as functions of $\eta$ for a $N=20$ system. At large negative values of $\eta$, which corresponds to constructive hybridization, the values of $\langle\bar{x}\rangle$ are negative for all the eigenmodes, indicating that they are localized on the left-most edge. This was illustrated in Fig. \ref{gFig6} for $\eta=-2$. As $\eta$ increases slowly from 0 to 1, a small proportion of the eigenmodes become localized at the right-most edge while the majority of the eigenmodes are still localized at the left-most edge. An example is shown in Fig. \ref{gFig9}c for $\eta=0.5$. At the special case of $\eta=1$, which was illustrated in Figs. \ref{gFig3} to \ref{gFig5}, the $\langle\bar{x}\rangle$ values of all the eigenmodes are 0, indicating that either every eigenmode is symmetrically localized both at the left and right edges as in Fig. \ref{gFig4} or \ref{gFig5}, or that it is a symmetrical bulk state, as in Fig. \ref{gFig6} (which can be interpreted as a state localized at both the left and right edges when the localization length exceeds half the length of the system). 

As $\eta$ increases slightly above 1, the distributions of some of the eigenstates start to shift towards the left edge. The number of modes exhibiting this distribution shift towards the left edge and the extent of the shift increase with $\eta$, resulting in an increasing number of states having a net localization at the left edge. An example of this is shown for $\eta=1.5$ in Fig. \ref{gFig9}c, at which roughly a third of the eigenstates are localized at the left edge. As $\eta$ increases further, a critical value is reached at which half the states are localized at the left edge (negative $\langle\bar{x}\rangle$ and half the states on the right edge (positive $\langle\bar{x}\rangle$). The critical value of $\eta$ increases slightly with $\delta_0$.  The eigenstates remain equally distributed between states localized at the left edge and states localized at the right edge as $\eta$ is further increased. An example of this ``half-half skin localization" was shown in Fig. \ref{gFig7}. 

\subsection{Mechanism for constructive and destructive hybridization}
To explain the mechanism behind the constructive and destructive hybridization, we plot the loci traced out by the eigenvalues of the surrogate Hamiltonian $H_{\mathrm{two-chains}}(\beta)$ for the coupled chain system obtained by substituting $k_x = -i\ln(\beta)$ into Eq. \eqref{reeq1} as functions of $|\beta|$ when $\mathrm{arg}(\beta)$ is varied from $-\pi$ to $\pi$. For a given eigenenergy $E$, there are eight values of $\beta$ that satisfy the Schr\"{o}dinger equation $H_{\mathrm{two-chains}} (\beta)|\psi_{\beta;E}\rangle = |\psi_{\beta;E}\rangle E$ for some $|\psi_{\beta;E}\rangle$. Two of the $\beta$s are 0,  another two are $\infty$, and the remaining four have finite values (see the Appendix for details). We label the $\beta$ values (including $\beta=0$ and $\beta=\infty$) as $\beta_1$ to $\beta_8$ where $|\beta_1| \leq |\beta_2| \hdots \leq |\beta_8|$. An $E$ appears in the OBC spectrum in the thermodynamic limit only when the moduli of its middle $\beta$ values match, i.e., when $|\beta_4|=|\beta_5|$. In contrast, for each isolated chain, there are only two $\beta$ values for a given eigenenergy $E$.  The energy GBZ of an isolated chain is therefore the locus of $E$ where the moduli of the two $\beta$ values coincide with each other.   

\begin{figure}[ht!]
\centering
\includegraphics[width=0.9\textwidth]{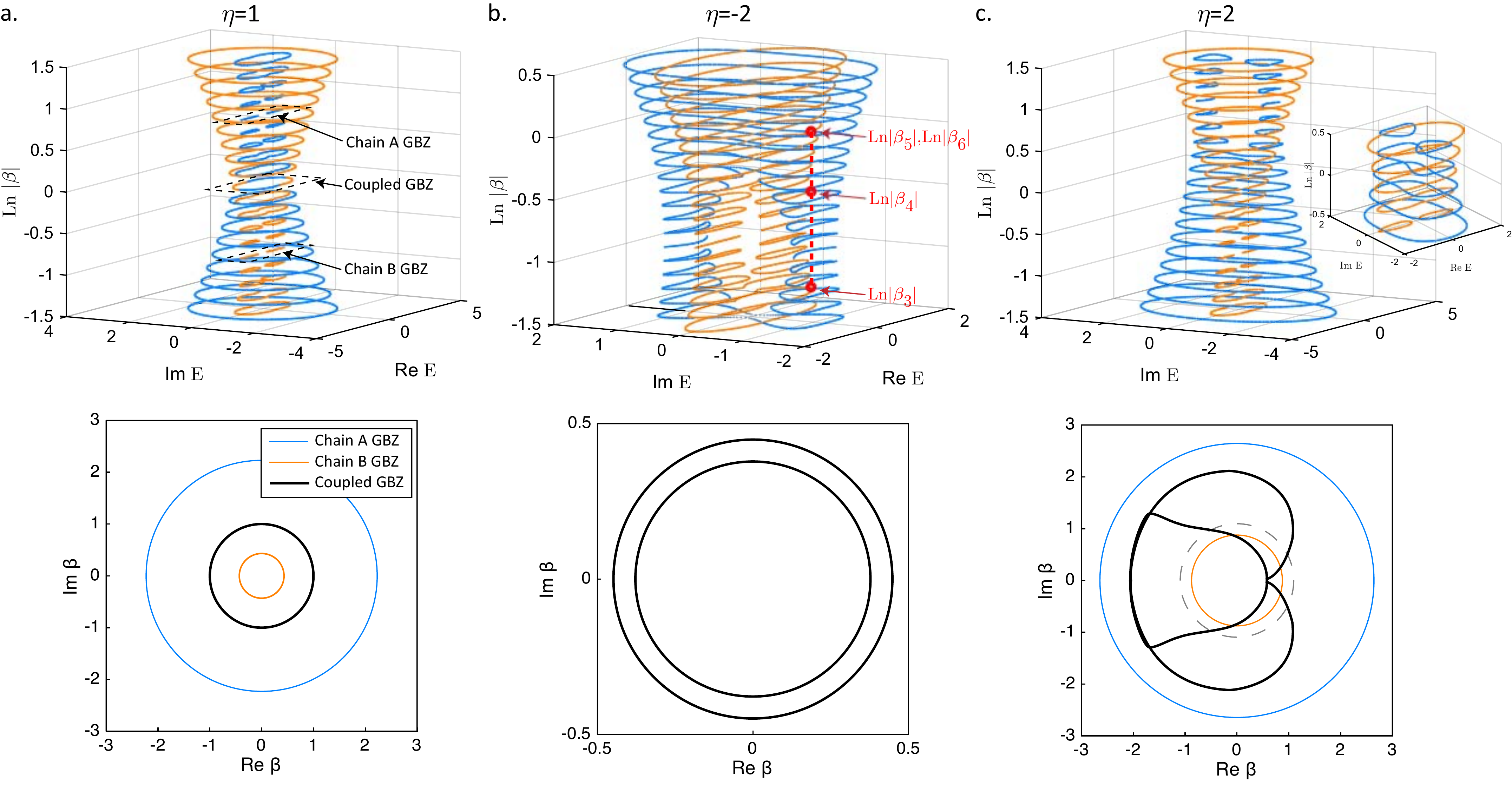}
\caption{ \textbf{Construction of GBZ under perfectly destructive, partially destructive, and constructive hybridization.} The upper panels show the loci spanned by the eigenenergies of $H_{\mathrm{two-chains}}(\beta)$ at different values of $|\beta|$ as $\mathrm{arg}\ \beta$ is varied between $-\pi$ to $\pi$ for \textbf{a} the perfectly destructive hybridization case in Fig. \ref{gFig3} ($\eta=1$), \textbf{b} constructive hybridization case in Fig. \ref{gFig6} ($\eta=-2$), and \textbf{c} partially destructive hybridization in Fig. \ref{gFig7} ($\eta=2$). The eigenenergies are color-coded according to which chain their corresponding eigenstates are mainly localized on.  The vertical dotted red line in panel b represents one constant value of complex energy. The red circles on the line denotes the values of $\ln|\beta|$ at which this energy is an eigenenergy of $H_{\mathrm{two-chains}}(\beta)$. The labels $\beta_3,\ldots\beta_6$ indicate which values of $\beta$ these eigenenergies correspond to. The inset in panel c is a zoomed in view of the eigengenergy loci near $\ln|\beta|=0$ showing that the eigenenergies from two different states coincide across a range of $|\beta|$ values. The lower panels show the corresponding $\beta$ GBZs of the isolated chains and the coupled system. The GBZ of the coupled system and isolated chains coincide in panel b. The grey dotted circle in panel c denotes the unit circle in the complex plane. Common parameters used: $t_1=1.2$,  $t_2= 0.6$, $\gamma_1= 0.8$, and $\delta_0 = 0.2$ } 
\label{gConVsDestIntf} 
\end{figure}

The fact that two different $\beta$ values at an eigenenergy $E$ have the same moduli $|\beta|$ implies that at that value of $|\beta|$, there are two values of eigenenergy, which are distinct from each other in the vicinity of $|\beta|$, coincide with each other at $|\beta|$. This can be seen, for example, in the upper plot of Fig. \ref{gConVsDestIntf}a in which the eigenenergies are color-coded according to whether they are localized mostly on chain A or chain B. The eigenenergy loci for the states localized in each chain takes the rough form of two cones with axes parallel to the $|\beta|$ axis where the apex of the upper (lower) cone points downwards (upwards). The cones are connected together at the apexes, near which the cones taper into the form of lines largely parallel to the $\mathrm{Re}\ E$ axis. Because of the small value of $\delta_0 = 10^{-7}$, the eigenenergy spectrum of the pair of coupled chains is very close to that of the two isolated chains combined together. Ignoring the eigenenergies localized on Chain B for the moment, the eigenenergies localized on Chain A shown in Fig. \ref{gConVsDestIntf}a is a very good approximation of the eigenenergy loci of the isolated Chain A. Around $\mathrm{ln}|\beta|=0.8$, the eigenenergy loci for Chain A takes the form of two disconnected ellipses in which the minor axis becomes narrower until the ellipses collapse into the lines labeled ``Chain A GBZ'' in the figure. The collapse of the ellipses into lines implies that the eigenenergies on either side of the ellipse major axis coincide with one another. In other words, the lines represent the loci at which the eigenenergies have two coincident values of $|\beta|$, and hence form the energy GBZ of the isolated chain A. Because the lines have the same value of $|\beta|$, the $\beta$ GBZ of the isolated Chain A is a circle of uniform radius on the complex plane, as shown in the lower plot of Fig. \ref{gConVsDestIntf}a.  Similarly, if we now ignore the eigenenergies of the states largely localized on Chain A and only look at those largely localized on Chain B, the two lines parallel to the $\mathrm{Re}\ E$ axis near $\mathrm{ln}|\beta|=-0.8$ are the energy GBZ of the isolated chain B, and give rise to a $\beta$ GBZ in the form of a circle on the complex plane with a smaller radius than that of the isolated chain A GBZ. 

We now consider the coupled system in Fig. \ref{gConVsDestIntf}a comprising both chains A and B. Although the energy GBZs of the individual isolated chains near $\mathrm{ln}|\beta|=\pm 0.8$ are loci of energy values where the $|\beta|$ values coincide, these coincident $\beta$ values are not the middle $|\beta|$ values $|\beta_4|$ and $|\beta_5|$ of the coupled chain. The GBZ of the isolated chain A (B) is where $|\beta_5|=|\beta_6|$ ($|\beta_3|=|\beta_4|$). The GBZ of the coupled system is the loci of the energies at which $|\beta_4|=|\beta_5|$, which occurs at $\ln|\beta|=0$ where the eigenenergy loci of chains A and B perfectly overlap each other. ( We provide an intuitive explanation of why the energy GBZ is constituted by the loci of the energy values at which the two middle $|\beta|$ values coincide in the Appendix.) Fig. \ref{gConVsDestIntf}a corresponds to the perfectly destructive hybridization case in Fig. \ref{gFig3} in which the two chains are antisymmetric with respect to each other. In this case, the eigenenergy loci of the states largely localized on the two chains are identical to each other except that they are symmetrically displaced with respect to each other about the $\mathrm{ln}|\beta|=0$ plane. The latter can be seen from the fact that the GBZs of the isolated chains are located at the same magnitude but opposite signs of $\mathrm{ln}|\beta|=\kappa$ (recall that $\kappa$ is the inverse skin length). The hybridization is therefore `destructive' because the opposite signs of $\kappa$ for the two isolated chains, which correspond to either $|\beta_3|=|\beta_4|$ or $|\beta_5|=|\beta_6|$, would mean that the GBZ of the coupled system, which corresponds to $|\beta_4|=|\beta_5|$, would be located at some intermediate value of $\kappa=\ln|\beta|$ with a smaller value of $|\kappa|$ than that of either isolated chain. The hybridization is `perfectly destructive' here because $\kappa=0$. 

Fig. \ref{gConVsDestIntf}b shows the eigenenergies as functions of $\ln|\beta|$ for the constructive hybridization case shown in Fig. \ref{gFig6} where the $\kappa = \ln|\beta|$ of each of the individual uncoupled chains have the same signs. Similar to the perfectly destructive case in Fig.  \ref{gConVsDestIntf}b, the energy GBZs of the individual uncoupled chains correspond to those values of $\ln|\beta|$ at which the loci of the eigenergies collapse from ellipses into lines. In contrast to Fig. \ref{gConVsDestIntf}b, however, the fact that $\kappa$ values of the two uncoupled chains have the same sign leads to these $\kappa$ values being at the same time the middle $\kappa$ values of the coupled system (i.e., the logarithms of $|\beta_4|=|\beta_5|$ at their respective energy values). The GBZ of the coupled system in this case is simply the union of the GBZs of the individual uncoupled chains, as is also indicated in the $\beta$ GBZ in the lower plot in Fig. \ref{gConVsDestIntf}b. The hybridization between the two chains is therefore `constructive' because the $\kappa$ values of the two individual isolated chains all have the same sign. Note that although there are values of $|\beta|$ besides those corresponding to the GBZs of the isolated chains at which the energies from two different chains coincide, these points of coincidence do not correspond to the middle $|\beta|$ values of the respective energies, which therefore do not appear on the energy GBZ of the coupled system. One exemplary value of complex energy at which the energies of the states localized mainly on two different chains coincide is indicated by the red vertical dotted line in Fig. Fig. \ref{gConVsDestIntf}b. It can be seen that point of coincidence occurs at $|\beta|=|\beta_5|=|\beta_6|$ because the other two values of $\beta$ (i.e., $\beta_3$ and $\beta_4$) which correspond to the same energy have smaller magnitudes.  ($\ln|\beta_1|=\ln|\beta_2|=-\infty$ and $\ln|\beta_7|=\ln|\beta_8|=+\infty$ are not shown in the plots.) 

The upper plot of Fig. \ref{gConVsDestIntf}c shows the eigenenergies as functions of $\mathrm{ln}|\beta|$ for the partially destructive hybridization case shown in Fig. \ref{gFig7} where the $\kappa$s of the individual isolated chains have opposite signs and different magnitudes. Unlike the perfectly destructive interference case shown in Fig. \ref{gConVsDestIntf}a, the eigenenergy loci of the states localized mainly on each chain are no longer identical to each other. Moreover, the GBZ of the individual isolated chains now have different values of $|\kappa|$. As a result, the the eigenenergies of the states mainly localized on the two different chains no longer all coincide only at $\ln|\beta|=0$ but rather, coincide across a range of $|\beta|$ values, part of which are shown in the inset of the upper plot. Therefore, the $\beta$ GBZ of the coupled system is no longer a constant-radius circle on the complex plane but instead has a more complicated form, as shown in the lower plot in Fig. \ref{gConVsDestIntf}c. For this particular set of parameters, the values of $|\beta|$ at which the eigenenergies of two different states coincide span across both sides of $\kappa=0$. This is reflected in the fact that part of the $\beta$ GBZ of the coupled system lying inside the complex unit circle shown in the grey dotted line and part of the $\beta$ GBZ lying outside the unit circle in the lower plot.  Some of the eigenstates of the finite-length coupled system are therefore localized near one edge of the system, and the remainder on the opposite edge in the half-half skin localization shown in Fig. \ref{gFig7}.

\section{Conclusion}
In this work, we have shown how the inter-chain coupling can be used for highly tunable NHSE in a coupled non-Hermitian lattice. We analyzed a concrete setup  consisting of two copies of the non-Hermitian  SSH chains with dissimilar skin solutions connected by inter-chain hopping. The skin eigenstates grow exponentially at both ends of short open systems with very small inter-chain coupling. The eigenvalue spectra exhibits a discontinuous transition from a real to complex spectra when the system size increases beyond a critical length, which decreases with increasing of the inter-chain strength.  

Our results also show that in addition to the tuning of the CNHSE through the modulation of the inter-chain coupling strength, the localization of the eigenstates can be tuned between constructive and destructive hybridization through the relative values of the non-Hermitian couplings between the two chains. For the perfectly destructive interference case, the OBC and PBC spectra do not vary with the system size, and at small inter-change couplings, every eigenstate exhibits an exponential decay from both edges of the system. A peculiar half-half skin localization  occurs under partially destructive interference in which some of the skin modes pile up exponentially on one end of a coupled system and the remainder on the opposite end. 

Our model highlights the effect and usefulness of multi-chain coupled non-Hermitian systems that go beyond the typical skin modes evolution and eigenvalue spectra of the single non-Hermitian chain. Furthermore, our works can easily be implemented in a variety of topolectrical circuit \cite{li2020critical,liu2020helical,yokomizo2021scaling} and optical platforms \cite{pal2017observing,schmidt2021coupled,perczel2020topological}. 

\subsubsection*{Acknowledgements}
This work is supported by the Ministry of Education (MOE) Tier-II grant MOE2018-T2-2-117 (NUS Grant Nos. R-263-000-E45-112/R-398-000-092-112), MOE Tier-I FRC grant (NUS Grant No. R-263-000-D66-114), \Rafi{MOE Tier-I grant R-144-000-435-133}, and other MOE grants (NUS Grant Nos. C-261-000-207-532, and C-261-000-777-532).
\bibliographystyle{apsrev4-1}

\section{Appendix : Critical non-Hermitian skin effect (CNHSE)} 
\subsection{Short-chain regime} 
	To explain the CNHSE, we first consider a simpler system consisting of two antisymetrically coupled Nelson-Hatano (NH) chains denoted as chains A and B for ease of explanation. Compared to the non-Hermitian SSH chain  in the main body, the NH chain is a simpler system because the unit cell of each NH chain contains only a single node rather than the two nodes in each SSH chain. The forward and backwards couplings between a node and its neighbor on the same chain in the chain A (B) are denoted as $t_2-\gamma$  ($t_2+\gamma$) and $t_2 - \gamma$ ($t_2+\gamma$), respectively, and the inter-chain coupling as $\delta_0$.  We thus write the surrogate Hamiltonian for the antisymetrically coupled NH chain system as 
\begin{equation}
	H_{\mathrm{NH}}(\beta) = \begin{pmatrix}  (t_2 + \gamma)\beta + (t_2 - \gamma)\beta^{-1} & \delta_0 \\ \delta_0 & (t_2 - \gamma)\beta + (t_2 + \gamma)\beta^{-1} \end{pmatrix}. \label{Hnh}
\end{equation}

We first consider $\delta_0 = 0$, the chains are uncoupled, and the OBC eigenenergies of the uncoupled chains fall on the real line, similar to what we have earlier seen for the uncoupled SHH chains in Fig. \ref{gFig2} . The condition for a given energy value of energy to appear in the OBC spectrum of an isolated chain is that the two $\beta$ at that energy should have the same moduli. For a given real eigenvalue $E$ that appears on the OBC spectrum of the isolated chain, there are therefore two eigenstates of Eq. \eqref{Hnh} with $\beta$ values having the same magnitude but different phases with the eigenspinor $(1,0)^\mathrm{T}$, which corresponds to states localized on chain A, and two eigenspinors with another two different values of $\beta$ with the eigensinor $(0, 1)^\mathrm{T}$, which correspond to states localized on chain B respectively. 

Let us now turn on the interchain coupling $\delta_0$ to a small finite value. To linear order in $\delta_0$, the two $(1, 0)^\mathrm{T}$ eigenstates for the same eigenenergy $E$ now become $|\tilde{\mathrm{A}}_j\rangle = (1, \delta_0 a_j)^\mathrm{T}$ where $a_j$ is independent of $\delta_0$. We denote the exact normalized form of $|\tilde{\mathrm{A}}_j\rangle$ as $|\mathrm{A}_j\rangle$. We find numerically that in the small $\delta_0$, short-chain regime where the OBC spectrum of the weakly coupled chains resembles that of the isolated chains, the states that are largely localized on the same chain have the same values of $|\beta|$. We therefore write the $\beta$ values for $|\mathrm{A}_j\rangle$and $|\mathrm{B}_j\rangle$ as 
\begin{equation}
	\beta_{\mathrm{A};j} = |\beta_{\mathrm{A}}|\exp(ik_{\mathrm{A};j}) \label{betaA}
\end{equation}
where $k_{\mathrm{A};j}$ is the real argument of  $\beta_{\mathrm{A};j}$. Similarly, we write the normalized eigenvectors of the states largely localized on chain B as $|\mathrm{B}_j\rangle$ with the corresponding $\beta$ values 
\begin{equation}
\beta_{\mathrm{B}; j} = |\beta_{\mathrm{B}}|\exp(i k_{\mathrm{B};j})
\label{betaB}
\end{equation}

Setting the origin $x=0$ at the center of the coupled chain system of length $N$, by symmetry and using Eqs. \eqref{betaA} and \eqref{betaB}, the wavefunction of an OBC eigenstate $\psi_{\mathrm{NH}}(x)$ in the small $\delta_0$, short-chain regime takes the form of 
\begin{equation}
	\psi_{\mathrm{NH}}(x) = \sum_{j=(1,2)} |\mathrm{A}_j\rangle |\beta_{\mathrm{A}}|^x \exp(i k_{\mathrm{A};j} x) \exp(i c_{\mathrm{A};j}) + |\mathrm{B}_j\rangle |\beta_{\mathrm{B}}|^x \exp(i k_{\mathrm{B};j} x) \exp(i c_{\mathrm{B};j}) \label{wfnh}
\end{equation}
where the $\exp(i c_{\mathrm{A/B};j})$s are the unit-modulus weights of the respective infinite-length eigenstates necessary to ensure that the coupled chain ensures the boundary condition $\psi(x=-N/2)=\psi(x=N/2) = (0,0)^\mathrm{T}$.  Consider now the boundary condition at the right edge $\psi(x=N/2) = (0,0)^\mathrm{T}$. We have
\begin{align}
	&\psi(X=N/2) = \begin{pmatrix} 0 \\ 0 \end{pmatrix} \nonumber \\
	\Rightarrow& |\beta_{\mathrm{A}}|^{N/2} \sum_{j=1,2} |\mathrm{A}_j\rangle \exp(i k_{\mathrm{A};j} N/2 )\exp(i c_{\mathrm{A};j}) = - |\beta_{\mathrm{B}}|^{N/2} \sum_{j=1,2} \exp(i c_{\mathrm{A};j}) \exp(i k_{\mathrm{B};j} N/2) \exp(i c_{\mathrm{B};j}).  \label{nhwl1} 
\end{align}

Taking the norm of both sides of the equal sign in Eq. \eqref{nhwl1} and introducing $\exp(i d_{\mathrm{A,B};j}) \equiv \exp(i c_{\mathrm{A,B}; j})\exp(i k_{\mathrm{A, B}; j} N/ 2)$, Eq. \eqref{nhwl1} implies that
\begin{equation} 
\Big| \sum_{j=1,2}|\mathrm{A}_j\rangle \exp(i d_{\mathrm{A};j}) \Big| = \left(\frac{|\beta_{\mathrm{B}}|}{|\beta_{\mathrm{A}}|}\right)^{N/2} \Big| \sum_{j=1,2}|\mathrm{B}_j\rangle \exp(i d_{\mathrm{B};j}) \Big |. \label{nhwl2} 
\end{equation}
For a finite $\delta_0$, the terms on both sides of the equal sign in Eq. \eqref{nhwl2} satisfy
\begin{align}
	&0 < \Big| \sum_{j=(1,2)}|\mathrm{A}_j\rangle \exp(i d_{\mathrm{A};j}) \Big| < 2 \\
	&0 < \Big| \sum_{j=(1,2)}|\mathrm{B}_j\rangle \exp(i d_{\mathrm{B};j}) \Big| < 2 
\end{align}
because the $|\mathrm{A/B}_j\rangle$s are normalized while $|\exp(i d_{\mathrm{A/B};j})| = 1$. This in turn implies that there is a maximum value of $N$ for which Eq. \eqref{nhwl2} can be satisfied. For example, if $|\beta_{\mathrm{B}}/\beta_{\mathrm{A}}| > 1$, the largest possible value of $N$ at which Eq. \eqref{nhwl2} can possibly be satisfied occurs at the largest possible value of $\big| |\mathrm{A}_1\rangle + |\mathrm{A}_2\rangle\exp(i f) \big|$ and the smallest possible value of $\big| |\mathrm{B}_1\rangle + |\mathrm{B}_2\rangle\exp(i g) \big|$ as $f$ and $g$ vary from $-\pi$ to $\pi$. We denote these two values as $\nu_{\mathrm{max}}$ and $\nu_{\mathrm{min}}$, respectively, so that at the largest possible value of $N=N_{\mathrm{max}}$, we have 
\begin{align} 
	&\nu_{\mathrm{max}} = |\beta_{\mathrm{B}}/\beta_{\mathrm{A}}|^{N_{\mathrm{max}}} \nu_{\mathrm{min}} \nonumber \\
	\Rightarrow& N_{\mathrm{max}} = (\ln(\nu_{\mathrm{max}}/\nu_{\mathrm{min}}))/(\ln |\beta_{\mathrm{B}}/\beta_{\mathrm{A}}|) 
\end{align} 

\begin{figure}[ht!]
\centering
\includegraphics[width=0.7\textwidth]{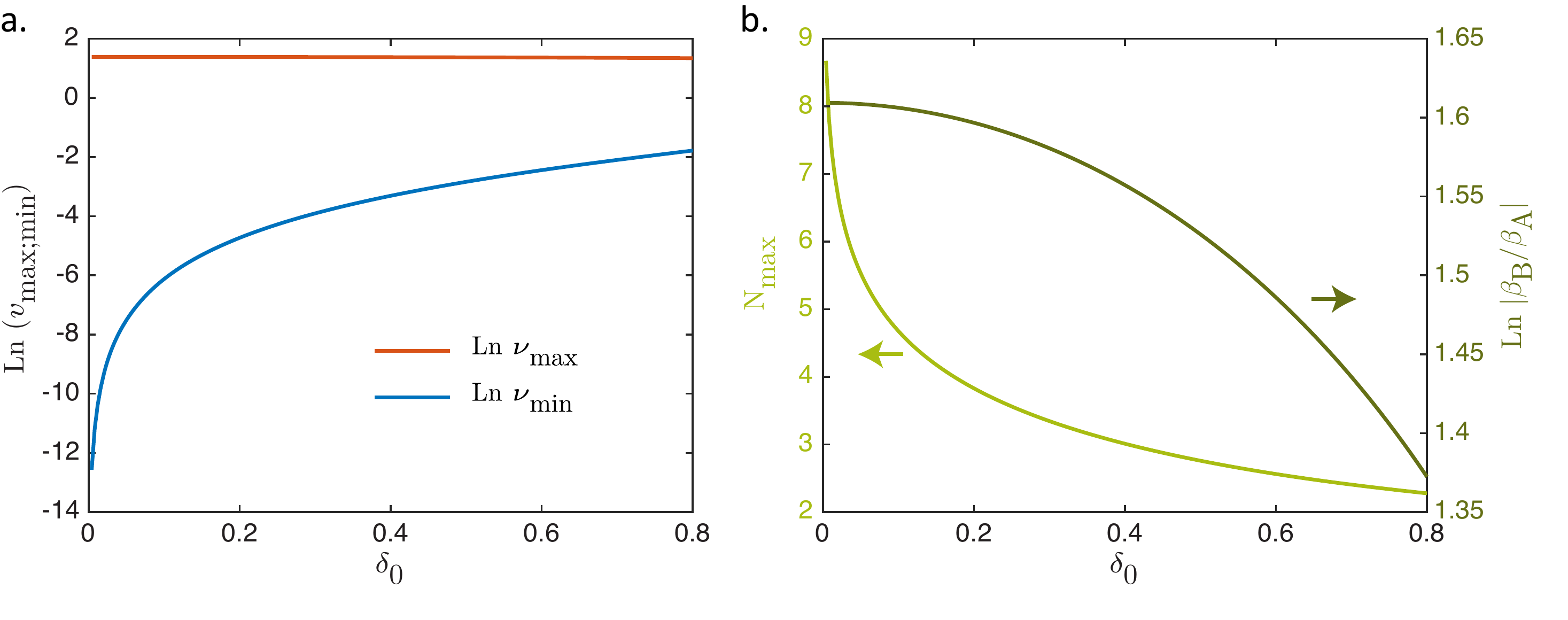}
\caption{ \textbf{Decrease in maximum chain length for the short-chain regime $N_{\mathrm{max}}$ before emergence of CNHSE with increasing inter-chain coupling strength $\delta_0$.}  a. $\ln(\nu_{\mathrm{min,max}})$ for $E=1$, $t_2=1.2$, and $\gamma=0.8$. b. The corresponding $N_{\mathrm{max}}$ and $\ln |\beta_{\mathrm{B}}/\beta_{\mathrm{A}}|$. } 
\label{gAppNHnum} 
\end{figure}

Fig. \ref{gAppNHnum} shows an exemplary plot of $\ln(\nu_{\mathrm{min/max}})$, $\ln |\beta_{\mathrm{B}}/\beta_{\mathrm{A}}|$, and $N_{\mathrm{max}}$ for one particular value of $E$ that falls on the energy GBZ of the isolated chain for an exemplary set of parameters. Increasing the inter-chain coupling $\delta_0$ reduces the difference between the magnitudes of $\nu_{\mathrm{max}}$ and $\nu_{\mathrm{min}}$ (panel a). Although the magnitude of $\ln|\beta_{\mathrm{B}}/\beta_{\mathrm{A}}|$ is also reduced slightly with increasing $\delta_0$, the reduction is relatively small. $N_{\mathrm{max}}$, the largest possible value of $N$ at which the wavefunction may still take the form of Eq. \eqref{wfnh} and the OBC energy spectrum of the finite-length coupled chain system still resembles that of energy GBZ of the uncoupled chains,  therefore decreases with $\delta_0$. When the length exceeds $N_{\mathrm{max}}$, Eq. \ref{wfnh} can no longer satisfy the boundary conditions at the edges. This results in a change in form of the wavefunction and consequently, the CNHSE, in which the energy GBZ of the coupled chain system now begins to approach its form in the thermodynamic limit. 

\subsection{Longer chains}  
We find numerically that in the intermediate-length regime where $N$ is still finite but longer than $N_{\mathrm{max}}$, the constituent periodic eigenstates that constitute the finite length eigenstates now have different $|\beta|$s values. Let us now label the periodic eigenstates as $|v_j\rangle$, $j\in(1,2,3,4)$ with corresponding $\beta$ values $\beta_j=|\beta_j|\exp(i k_j)$s and weights $c_j$. The eigenstates are labeled so that $|\beta_1|<|\beta_2|<|\beta_3|<|\beta_4|$. 
Again setting $x=0$ in the middle of the chain, the wavefunction of the finite length system now takes the form of 
\begin{equation}
	\psi_{\mathrm{NH}}(x) = \sum_{j=1,2,3,4} |v_j\rangle |\beta_j|^x \exp(i k_j x)c_j \label{wfil1}.
\end{equation} 

Consider now the boundary condition at the right edge that $\psi_{\mathrm{NH}}(N/2) = (0,0)^{\mathrm{T}}$. Introducing $\varphi_j = k_jN/2 + \mathrm{arg} c_j  - (k_3N/2 + \mathrm{arg} c_3)$, we have
\begin{equation}
	|v_1\rangle\left( \frac{|\beta_1|}{|\beta_3|} \right)^{N/2}|c_1|\exp(i \varphi_1) +|v_2\rangle \left( \frac{|\beta_2|}{|\beta_3|} \right)^{N/2} |c_2|\exp(i \varphi_1) + |v_3\rangle |c_3| + |v_4\rangle \left(\frac{|\beta_4|}{|\beta_3|}\right)^{N/2}|c_4|\exp(i\varphi_4) = \begin{pmatrix} 0 \\ 0 \end{pmatrix}.  \label{wfnhr} 
\end{equation}

\begin{figure}[ht!]
\centering
\includegraphics[width=0.7\textwidth]{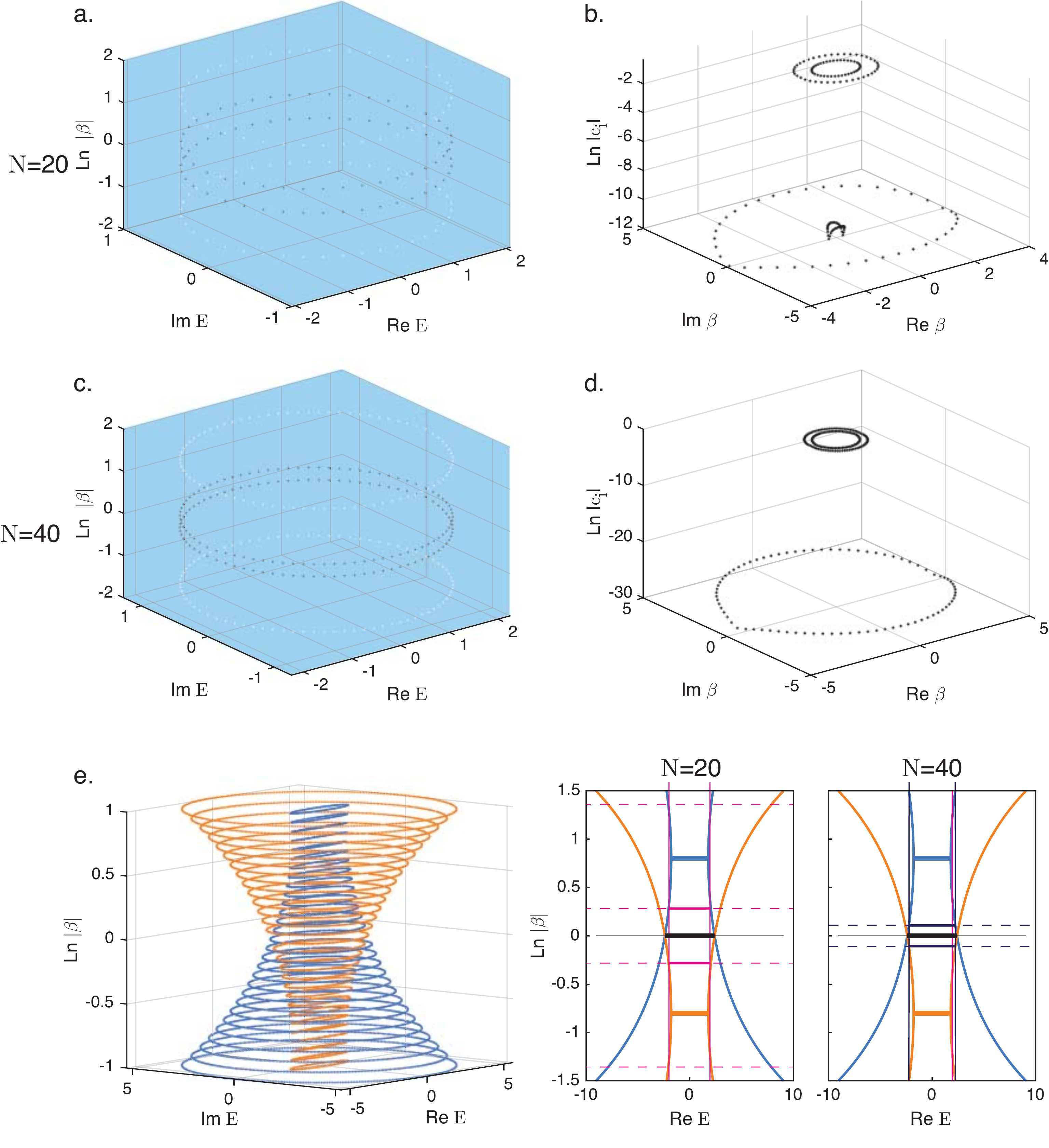}
\caption{ \textbf{Evolution of coupled chain eigenspectrum with increasing chain length.} \textbf{a} The $(x,y)$ coordinates of each point show the complex eigenenergies for a 20 unit cells-long coupled chain system with $t_2=1.2$,$\gamma=0.8$, and $t_0=5\times 10^{-3}$. The $z$ coordinate of the four points at each eigenenergy indicate the magnitudes of the $\beta$ values of the four $|\beta|$ PBC eigenstates that constitute each OBC eigenstate. A darker color of each dot indicates a higher weightage $|c_i|$ for the dot. \textbf{b} The $(x, y)$ coordinates of each point indicates the complex $\beta$ values that are present in the finite-length OBC eigenstates of the 20 unit cells-long system in a. The $z$ coordinate of each $\beta$ value indicates its weight in the finite-length OBC eigenstate wavefunction.  \textbf{c, d} are the counterparts to a and b for a 40 unit cells-long coupled chain system. \textbf{e}. The left plot shows the loci of the eigenenergies of $H_{\mathrm{NH}}(\beta)$ as functions of $|\beta|$ as $\mathrm{arg}\ \beta$ is varied from $-\pi$ to $\pi$. The colors of the dots indicate the chain the corresponding eigenstates are primarily localized on.  The middle and right plots show the loci as functions of $\mathrm{Re}\ E$ axis at $\mathrm{Im}\ E = 0$. The thick horizontal lines at $\ln|\beta|=\pm 0.8$ are the OBC energy GBZ of the isolated chain, and that at $\ln|\beta|=0$ is the projection of the OBC GBZ of the coupled-chain system in the thermodynamic limit. The thin vertical lines denote the values of the real eigenenergies of the (middle plot) $N=20$ system and (right plot) $N=40$ coupled-chain systems. The dotted horizontal lines indicate the $\ln|\beta|$ values that the PBC eigenstates at the eigenenergies of the coupled-chain systems correspond to. } 
\label{gnhFinLenComp} 
\end{figure}

The first term on the left of the equal sign proportional to $(|\beta_1/\beta_3|)^{N/2}$ becomes vanishingly small as $N\rightarrow\infty$ because, by construction $|\beta_1/\beta_3| < 1$. Its contribution to fulfilling the boundary condition at the right edge is therefore negligible and, to a first approximation, can be ignored. We next note that the $|v_i\rangle$s are linearly independent $2\times 1$ vectors. In order to satisfy the condition that the terms containing $|v_2\rangle$,$|v_3\rangle$, and $|v_4\rangle$ sum up to $(0,0)^\mathrm{T}$, it is required that their weights (i.e., $c_j\exp(ik_jN/2)$) should be roughly of the same order of magnitude. Let us set $c_3 = 1$ for convenience. The weight of the term containing $|v_2\rangle$ contains a factor of $|\beta_2/\beta_3|^{N/2}$. In order to ensure that $|\beta_2/\beta_3|^{N/2} \simeq 1$ regardless of how large $N$ grows, we need $|\beta_2|$ and $|\beta_3|$ to approach each other as $N$ grows towards infinity because $|c_2|=|c_3|$ by symmetry. This gives the well-known condition that the energies that appear on the OBC GBZ are those where the middle $|\beta|$ values match each other. Meanwhile, to ensure that the weight of the term containing $|v_4\rangle$ is roughly on the order of 1, we also require $(|\beta_4/\beta_3|^{N/2}|c_4|)\simeq 1$. Therefore, the value of $|c_4|$ decreases with the length of the system. It can be shown that the weight of $|c_1|$ similarly decreases with the length of the system by considering the the boundary condition at the left edge of the system at $x=-N/2$. These predictions are borne out by the explicit comparisons of the $|c_i|$s and $|\beta_i|$s shown in Fig. \ref{gnhFinLenComp} for the eigenstates of two systems with the same parameters as that in Fig. \ref{gAppNHnum} and $\delta_0 = 5 \times 10^{-3}$. 
 
Fig. \ref{gnhFinLenComp}a and c show the energies of eigenstates of the 20 unit cell-long (panel a) and 40 unit cell-long (panel c) coupled chains, and for each eigenstate, the values of the $|\beta_i|$s constituting the eigenstate, and the relative weightages (i.e., $|c_i|$s) of each $|\beta_i|$ value through how dark each point is. In line with our expectation, the weights of the middle $|\beta|$ values dominate over those of the smallest and largest $|\beta|$ values. Comparing panels a and c, it is evident that as expected, the middle $|\beta|$ values approach each other as the length of the system increases. 

Fig. \ref{gnhFinLenComp}b and d show the collection of all the $|\beta|$ values that are present in all finite-length eigenstates and their weightages in their respective eigenstates. (The $|\beta_1|$ values are missing in panel d because their magnitudes were smaller than the machine precision and were approximated to 9 by the numerical algorithm used to decompose the finite-length eigenstate into its constituent $|v_i\rangle$s.) In agreement with the panels a and c, the components with large weights are the middle $|\beta|$s, which constitute the two rings with the second ($\beta_3$s) and third largest ($\beta_2$s) diameters. 

Comparing between panels b and d, it is evident that as expected, the $|\beta_2|$s tend to approach the $|\beta_3|$s as the length of the coupled system increases, while the weights of the $\beta_1$s and $\beta_4$ (i.e. the rings with the smallest and largest diameter, respectively) decrease as the length of the system increases. The left plot in Fig. \ref{gnhFinLenComp}e shows the loci of the eigenenergies of $H_{\mathrm{NH}}(\beta)$ with the variation of $|\beta|$. The loci is very similar to the loci of the perfectly destructive hybridization case shown earlier in Fig. \ref{gConVsDestIntf}a. The middle and right plots show the loci at as functions of $\mathrm{Re} E$ at $\mathrm{Im}\ E = 0$ together with the OBC GBZ of the isolated chains and the projection of the OBC GBZ of the coupled chain system in the thermodynamic limit. The $|\beta|$ values of the PBC states making up the OBC eigenstates of the $N=20$ (midddle) and $N=40$ (right) chains are also marked out in the plots. Comparing the middle and right panels, it is evident that as the $|\beta_2|$ and $|\beta_3|$ values approach each other as the length of the coupled-chain system increases so that eventually $|\beta_2|=|\beta_3|=1$ in the thermodynamic limit, the corresponding magnitudes of the energies increase. This explains why the area enclosed by the energy spectrum on the complex energy plane increases with the size of the system in panel d of Fig. \ref{gFig3} and \ref{gFig4} after the CNHSE.  

\subsection{Critically hybridized HN chains vs SSH chains} 
The discussion in this appendix has so far focused on the coupled-NH chain system, which is a simpler system represented by $2\times 2$ Hamiltonians than the coupled-SSH chain system in the main body of the text represented by $4\times 4$ Hamiltonians to more clearly bring out the main features of CNHSE and the accompanying transition from the short-chain regime to the long-chain regime. Here, we briefly outline the main differences between the coupled NH and the coupled SSH chain systems. 

The surrogate Hamiltonian for a generic system with nearest unit-cell coupling, such as the coupled NH and SSH chain systems we have studied so far, can generically be written as $H(\beta) = H_+\beta + H_-\beta^{-1} + H_0 $ where $H_\pm$ are the $\beta$-independent parts of $H$ that multiply $\beta^{\pm 1}$, and $H_0$ is the part of $H(\beta)$ independent of $\beta$. For a given eigenstate $|\psi\rangle$ with eigenenergy $E$, the time-independent Schr\"{o}dinger   equation  $H|\psi\rangle= |\psi\rangle E$ can be cast into the form of 
\begin{equation}
	\begin{pmatrix} H_- & H_0 - E\mathbf{I}_d \\ \mathbf{0}_d & \mathbf{I}_d \end{pmatrix} \begin{pmatrix} |\psi\rangle/\beta \\ |\psi\rangle \end{pmatrix} = \beta \begin{pmatrix} \mathbf{0}_d & -H_+ \\ \mathbf{I}_d & \mathbf{0}_d \end{pmatrix}\begin{pmatrix} |\psi\rangle/\beta \\ |\psi\rangle \end{pmatrix} \label{gegEq}  
\end{equation} 
where $\mathbf{I}_d$ and $\mathbf{0}_d$ are respectively the identity and zero matrices with the same dimensionality as $H_0$ and $H_\pm$. Eq. \ref{gegEq} takes the form of a generalized eigenvalue equation 
\begin{equation}
	\mathbf{A}(E)\mathbf{v} = \beta\mathbf{B}\mathbf{v} \label{gegEq1}
\end{equation} 
 where $\mathbf{v}$ is $[|\psi\rangle/\beta; |\psi\rangle]^{\mathrm{T}}$, and $\mathbf{A}(E)$ and $\mathbf{B}$ are matrices twice the size of $H$ whose explicit forms can be read off from Eq. \eqref{gegEq}.  $\mathbf{A}(E)$ is an $E$-dependent matrix because of the $H_0 - E\mathbf{I}_d$ term. Given the eigenenergy $E$ and Hamiltonian $H$ so that $\mathbf{A}(E)$ and $\mathbf{B}$ can be constructed, Eq. \eqref{gegEq1} can be solved for $\beta$ and $\mathbf{v}$. For the coupled NH chain model discussed so far in the Appendix,  $\mathbf{A}$ and $\mathbf{B}$ are $4 \times 4$ matrices. There are therefore four solutions for $\beta$ for a given $E$. 

At the same time, the characteristic polynomial for $|H_{\mathrm{NH}}(\beta) - E\mathbf{I}_2|$ with the give value of $E$ can also be cast into a fourth-order polynomial in $\beta$. The four eigenvalues of the generalized eigenvalue problem in this case correspond to the four roots of $\beta$ in the characteristic equation in this case. In contrast, for the coupled-SSH chain system,  $\mathbf{A}$ and $\mathbf{B}$ are now $8\times 8$ matrices, so there are eight eigenvalues for $\beta$. However, the characteristic equation for $|H_{\mathrm{coupled}}(\beta) - E\mathbf{I}_4|$ is still a fourth-order polynomial with only four roots. The four `extra' eigenvalues besides the four roots of the characteristic equation consist of a pair of eigenvectors with the eigenvalue $\beta=0$, and another pair of eigenvectors with the eigenvalue $\beta=\infty$. These four eigenvalues and eigenvectors do not vary with any finite value of $E$ (including $E=0$).  The two $\beta=0$ eigenvectors, which we denote as $|v_1\rangle$ and $|v_2\rangle$ satisfy 
\begin{equation}
H_+|v_{1;2}\rangle = (0,0,0,0)^{\mathrm{T}}
\end{equation}
 and correspond to states localized at the right-most edge of a finite (or semi-infinte) system. The two $\beta=\infty$ eigenvectors, which we denote as $|v_7\rangle$ and $|v_8\rangle$, satisfy 
\begin{equation}
H_-|v_{7;8}\rangle = (0,0,0,0)^{\mathrm{T}}
\end{equation}
and correspond to states localized at the left-most edge of a finite system.

Let us first consider the short-chain, small-interchain coupling regime where the OBC energy spectrum of the coupled SSH system resembles that of the uncoupled system ($N \leq 19$ in Fig. \ref{gFig3}d and $N \leq 5$ in Fig. \ref{gFig4}d ). At small $\delta_0$, two of the four PBC eigenstates with finite $\beta$ values are localized mainly in Chain A , and we denote them as $|\mathrm{A}_{j}\rangle$  and their corresponding $\beta$ values as 
\begin{equation}
\beta_{\mathrm{A};j}=|\beta_\mathrm{A}|\exp(ik_{\mathrm{A};j})|
 , j\in(1,2).
\end{equation}
 The other two finite-$\beta$ states are localized on Chain B and their corresponding $\beta$ values and eigenvectors are analogously defined. The form of the wavefunction in the short-chain, smaller inter-chain coupling regime for the coupled SSH chains is almost the same as that for the coupled NH chains in Eq. \eqref{wfnh} except for the addition of the $\beta=0$ and $\beta=\infty$ states, and the fact that the eigenvectors are now $4\times 1$ vectors : 
\begin{align}
\psi_{\mathrm{two-chains}}(x) =& \sum_{j=1,2} \Big(  |\mathrm{A}_j\rangle |\beta_{\mathrm{A}}|^x \exp(i k_{\mathrm{A};j} x) \exp(i c_{\mathrm{A};j}) + |\mathrm{B}_j\rangle |\beta_{\mathrm{B}}|^x \exp(i k_{\mathrm{B};j} x) \exp(i c_{\mathrm{B};j}) \nonumber \\
&+ |v_j\rangle \delta_{x,-N/2} \gamma_j + |v_{j+6}\rangle \delta_{x, N/2} \gamma_{j+6} \Big) \label{wfssh1}
\end{align}
where the $\delta_{x, \pm N/2}$ are Kronecker deltas, the $\gamma_j$s are the coefficients of the $|v_j\rangle$ terms whose moduli is not necessarily one, and the remaining terms have similar meanings to the corresponding terms in Eq. \eqref{wfnh}. 

At the right boundary $x=L/2$, the terms containing $|v_1\rangle$ and $|v_2\rangle$ vanish, and the boundary condition $\psi(x=-L/2) = (0,0,0,0)^\mathrm{T}$ becomes 
\begin{equation}
 \sum_{j=1,2} \Big(   |\mathrm{A}_j\rangle |\beta_{\mathrm{A}}|^{N/2} \exp(i d_{\mathrm{A};j}) + |\mathrm{B}_j\rangle |\beta_{\mathrm{B}}|^{N/2} \exp(i d_{\mathrm{B};j} x) \Big) + \sum_{j=7,8} |v_j\rangle \gamma_j  = \begin{pmatrix} 0 \\ 0 \\ 0 \\ 0 \end{pmatrix} \label{wfssh2} 
\end{equation}
where $d_{\mathrm{A,B}; j} = k_{\mathrm{A,B};j}N/2 + c_{\mathrm{A,B};j}$. 
Recall that$|v_{1;2}\rangle$ satisfy $H_+|v_{1;2}\rangle = (0,0,0,0)^{\mathrm{T}}$. For the $H_+$ term contained within $H_{\mathrm{two-chains}}$ in Eq. \eqref{reeq1}, $|v_7\rangle$ and $|v_8\rangle$ can be set as $(1, 0, 0, 0)^\mathrm{T}$ and $(0,0,1,0)^\mathrm{T}$. Notice that the second and fourth vector components of $|v_7\rangle$ and $|v_8\rangle$ are both 0. Therefore, if we extract only the second and fourth vector components in both sides of the equal sign of Eq. \ref{wfssh2} terms and write down the resulting $2 \times 1$ vector equation, $\gamma_7$ and $\gamma_8$ would be absent from this equation. Writing the vectors formed by extracting the second and fourth components of $|\mathrm{A}_j\rangle$ and $|\mathrm{B}_j\rangle$ as $|\tilde{\mathrm{A}}_j\rangle$ and $|\tilde{\mathrm{B}}_j\rangle$, respectively, the vector equation formed by extracting the second and fourth components of Eq. \eqref{wfssh2} is
\begin{equation}
	|\beta_{\mathrm{A}}|^{N/2} \sum_{j=(1,2)} |\tilde{\mathrm{A}}_j\rangle \exp(i d_{\mathrm{A};j} N/2 ) = - |\beta_{\mathrm{B}}|^{N/2} \sum_{j=(1,2)} |\tilde{\mathrm{B}}_j\rangle \exp(i d_{\mathrm{B};j}) {\mathrm{B};j}). 
\end{equation}

This has exactly the same form as Eq. \eqref{nhwl1} with the exception that the norms of $|\tilde{\mathrm{A}},\tilde{\mathrm{B}}_j\rangle$ are now less than one because they contain only half the components of the normalized vectors $|\mathrm{A, B}\rangle$. Nonetheless, the argument for the emergence of the CNHSE in the coupled NH chain system that follows Eq. \eqref{nhwl1} can still be applied to the coupled SSH chain system here. 

Similarly, by repeating the trick of only focusing on the vector components which are 0 in both $|v_7\rangle$ and $|v_8\rangle$ in the corresponding equation to Eq. \eqref{wfnhr}, a $2\times 1$ vector equation with the same form as Eq. \eqref{wfnhr} can be obtained for the coupled SSH chain system. The same reasoning for why $|\beta_4|$ and $|\beta_5|$ approach each other and the area spanned by the OBC energy spectrum on the complex energy plane increases with the length of the system following Eq. \eqref{wfnhr} can then be applied to the coupled SSH chain system as well. 

%

\end{document}